\documentclass{article}

\usepackage[dvips]{graphicx} 
\usepackage{amssymb} 
\usepackage{amsmath} 
\usepackage{color} 
\usepackage{epsfig} 

\newtheorem{lemma}{Lemma}

\newcommand{\la}{{\lambda}}
\newcommand{\La}{{\Lambda}}
\newcommand{\pd}{{\partial}}
\newcommand{\eqand}{{\quad \mathrm{and} \quad}}
\newcommand{\Veff}{{V_{\mathrm{eff}}}}
\newcommand{\Om}{{\Omega}}

\newcommand{\dSdx}{{\left(\frac{dS_x}{dx}\right)^2}}
\newcommand{\dSdy}{{\left(\frac{dS_y}{dy}\right)^2}}
\newcommand{\Hxy}{{H(x,y)}}
\newcommand{\Hyx}{{H(y,x)}}
\newcommand{\Hx}{{H(x)}}
\newcommand{\Hy}{{H(y)}}
\newcommand{\Axy}{{A(x,y)}}
\newcommand{\Ayx}{{A(y,x)}}
\newcommand{\Lxy}{{L(x,y)}}
\newcommand{\Gx}{{G(x)}}
\newcommand{\Gy}{{G(y)}}
\newcommand{\Omf}{{\Omega_\phi}}
\newcommand{\Omy}{{\Omega_\psi}}
\newcommand{\lag}{{\mathcal{L}}}
\newcommand{\ham}{{\mathcal{H}}}
\newcommand{\axb}{{\alpha(x)}}
\newcommand{\bxb}{{\beta(x)}}
\newcommand{\ayb}{{\alpha(y)}}
\newcommand{\byb}{{\beta(y)}}
\newcommand{\gx}{{\gamma(x)}}
\newcommand{\gy}{{\gamma(y)}}
\newcommand{\zy}{{\zeta(y)}}
\newcommand{\tzx}{{\xi(x)}}

\newcommand{\tx}{{\theta(x)}}
\newcommand{\cx}{{\chi(x)}}

\begin{document}
\title{Geodesics and Symmetries of Doubly-Spinning Black Rings}
\author{Mark Durkee\\
DAMTP, University of Cambridge\\
Centre for Mathematical Sciences\\
Wilberforce Road, Cambridge, CB3 0WA, UK\\
M.N.Durkee@damtp.cam.ac.uk}

\maketitle

\abstract{This paper studies various properties of the Pomeransky-Sen'kov doubly-spinning black ring spacetime.  I discuss the structure of the ergoregion, and then go on to demonstrate the separability of the Hamilton-Jacobi equation for null, zero energy geodesics, which exist in the ergoregion.  These geodesics are used to construct geometrically motivated coordinates that cover the black hole horizon.  Finally, I relate this weak form of separability to the existence of a conformal Killing tensor in a particular 4-dimensional spacetime obtained by Kaluza-Klein reduction, and show that a related conformal Killing-Yano tensor only exists in the singly-spinning case.}

\section{Introduction} \label{sec:intro}
In recent years there has been significant interest in general relativity in higher dimensions, in particular in the black hole solutions of both the vacuum Einstein equations and various supergravity theories.

It has become increasingly clear that the study of black holes is far more complicated in higher dimensions, even if we limit ourselves to globally asymptotically flat vacuum solutions.  This was most vividly demonstrated by the discovery \cite{ER:2001} of a black ring solution of the vacuum Einstein equations in five dimensions with horizon topology $S^1\times S^2$.  This shows that the higher-dimensional generalization of the black hole topology theorem \cite{HawkingEllis} is non-trivial.  Galloway and Schoen \cite{Galloway:2005} studied the allowed topologies in a general number of dimensions, and show that in five dimensions any connected sum of $S^1\times S^2$s and spaces homotopic to a 3-sphere is allowed.

Furthermore, there are black ring solutions where the mass and angular momenta $(M,J_1,J_2)$ of the spacetime coincide with allowed angular momenta for the more familiar Myers-Perry solutions \cite{Myers:1986}.  Thus black hole uniqueness, at least in the familiar sense, also fails here.  More recent work (eg \cite{Elvang:2007bi}, \cite{Elvang:2007sat}, \cite{Iguchi:2007}, \cite{Evslin:2007}) on solutions with multiple black objects shows that uniqueness is violated to a much greater degree, in fact continuously.  There is a detailed recent review of this field \cite{ER:2008}.

In this paper we study the properties of a particular solution to the five-dimensional vacuum Einstein equations, namely the doubly-spinning black ring.  This solution was constructed by Pomeransky and Sen'kov \cite{Pomeransky} using inverse scattering techniques.  It is a generalisation of the original Emparan-Reall black ring with rotation around the $S^2$ as well as the $S^1$.  The solution is rather more complicated than the singly spinning black ring (which it reduces to in a particular limit).  Kunduri, Lucietti and Reall \cite{Kunduri:2007} studied the extremal limit and near-horizon geometry of this solution, while Elvang and Rodriguez \cite{Elvang:2007bi} studied its phase structure, asymptotics and horizon.  There is a more general version of the solution, corresponding to an `unbalanced' ring with conical singularities, which is explicitly presented in \cite{Morisawa}.  The current literature on this spacetime is reviewed in \cite{ER:2008}, we give some brief details of its properties in Section \ref{sec:metric}.

We then move on to describe some newly discovered properties of the metric.  In summary, we show that: \begin{itemize}
\item The topology of the ergosurface changes as the black ring parameters vary.  For a ring with sufficiently small rotation about the $S^2$, the topology is $S^1\times S^2$, as in the singly spinning case.  However, for more rapid $S^2$ rotation the ergosurface has topology $S^3 \cup S^3$, consisting of a small sphere around the centre of the ring, and a large sphere enclosing the entire ring.  In the bounding case, where a topology change occurs, the surface `pinches' on an $S^1$. (Section \ref{sec:ergo}).
\item The Hamilton-Jacobi equation can be reduced to an ODE when we restrict it to fixed point sets of one of the axial Killing vector fields, corresponding to the `axis' and `equatorial plane' of the ring.  Qualitative properties of geodesics lying within these 2-surfaces can therefore be studied analytically using effective potential techniques. (Section \ref{sec:axisgeo})
\item In general, for both the singly-spinning and doubly-spinning cases, the Hamilton-Jacobi equation admits separable solutions in the case of null, zero energy geodesics. (Section \ref{sec:hj})
\item Null, zero energy geodesics exist inside the ergoregion, and correspond to lightlike particles coming out of the white hole horizon in the past, and falling into the black hole horizon in finite parameter time in the future. (Section \ref{sec:ergogeo})
\item The existence of these geodesics allows us to construct new coordinate systems for the black ring that are valid across the horizon (Section \ref{sec:coord}). \begin{itemize} \item In the singly spinning case, it is possible to construct a set of coordinates $(v,x,y,\tilde{\phi},\tilde{\psi})$ such that $v$, $\tilde{\phi}$, $\tilde{\psi}$ are constant along one of these geodesics.  These coordinate systems are regular at the future black hole horizon, and a particular subset of them cover the entire horizon.  The coordinate change given in \cite{ER:2006} is included in this family of coordinate systems, and hence this allows us to understand its geometric significance.
\item In the doubly-spinning case, the best approach is to construct coordinates $(v,x,y,\tilde{\phi},\tilde{\psi})$ where only $\tilde{\phi}$ and $\tilde{\psi}$ are constant along the geodesics, and a change of coordinate $v$ is made that simply makes the metric regular at the horizon (rather than demanding that it is constant along the geodesics).  Using this approach, we are able to present explictly a form for the doubly-spinning metric that is valid across the horizon, which has not been done in the literature to date. \end{itemize}
\item The null, zero energy separability of the Hamilton-Jacobi equation is related to the existence of a conformal Killing tensor in a 4-dimensional spacetime obtained by a spacelike Kaluza-Klein reduction of the black ring spacetime in the ergoregion, reducing along the asymptotically timelike Killing vector field. (Section \ref{sec:sym})
\item A pair of conformal Killing-Yano tensors exist for the 4-dimensional spacetime if, and only if, the associated ring is singly-spinning. (Section \ref{sec:cky})
\item The Klein-Gordon equation is not separable in ring-like coordinates, even if we restrict to looking for massless, time-independent solutions. (Section \ref{sec:kg})\end{itemize}

Finally, we then briefly discuss some possible extensions of this work, and its relation to other results.
\paragraph{Note added:}After the work of Section \ref{sec:ergo} was mostly complete, a similar analysis of the doubly spinning black ring ergoregion \cite{Elvang:2008erg} appeared, as part of a paper discussing the properties of ergoregions in various higher-dimensional solutions.  This includes a more detailed analysis of how the ergoregion merger occurs.

\section{The Doubly-Spinning Black Ring Spacetime} \label{sec:metric}
Here, we briefly describe some properties of the doubly-spinning black ring spacetime, in order to set up notation, and gather together some results that will be useful in what follows.  We also explore some interesting properties of the ergoregion, which will be relevant later when we move on to consider geodesics.

\subsection{Form of the metric} \label{sec:met}
The doubly rotating ring solution can be written in the form \begin{eqnarray} ds^2 &=&
  -\frac{\Hyx}{\Hxy}(dt+\Omega)^2 + \nonumber \\ && \frac{R^2 \Hxy}{(x-y)^2 (1-\nu)^2} \left[
\frac{dx^2}{\Gx}-\frac{dy^2}{\Gy}+\frac{\Ayx d\phi^2-2\Lxy d\phi d\psi- \Axy d\psi^2}{\Hxy \Hyx}\right].\label{eqn:metric}\end{eqnarray} The coordinates lie in ranges $-\infty<y\leq -1$, $-1\leq x \leq 1$ and $-\infty < t < \infty$, with $\psi$ and $\phi$ $2\pi$-periodic.  Varying the parameters $\la$ and $\mu$ changes the shape, mass and angular momentum of the ring.  They are required to lie in the ranges $0\leq\nu<1$ and $2\sqrt{\nu}\leq \la < 1+\nu$

The functions $G$, $H$, $A$ and $L$ are moderately complicated polynomials, and are
given by \begin{eqnarray*} \Gx&=&(1-x^2)(1+\la x + \nu x^2),\\
\Hxy&=& 1+\la^2-\nu^2+2 \la \nu(1-x^2)y+2 x \la(1-y^2 \nu^2)+x^2 y^2 \nu (1-\la^2-\nu^2),\\
\Lxy&=& \la \sqrt{\nu} (x-y)(1-x^2)(1-y^2) \left[1+\la^2-\nu^2+2(x+y)\la \nu - xy\nu(1-\la^2-\nu^2)\right],\\
\Axy&=& \Gx(1-y^2)\left[((1-\nu)^2-\la^2)(1+\nu)+y\la(1-\la^2+2\nu-3\nu^2)\right] \\
& & + \Gy\left[ 2\la^2+x \la((1-\nu)^2+\la^2)+x^2((1-\nu)^2-\la^2)(1+\nu) \right. \\
& & \left. + x^3
\la(1-\la^2-3\nu^2+2\nu^3)+x^4\nu(1-\nu)(1-\la^2-\nu^2)\right]\end{eqnarray*}  The rotation is described by the 1-form $\Om = \Om_\psi d\psi + \Om_\phi d\phi$, where \[ \Om_\psi = -\frac{R\la\sqrt{2((1+\nu)^2-\la^2)}}{\Hyx} \frac{1+y}{1-\la+\nu}\left(1+\la-\nu+x^2 y \nu (1-\la-\nu)+2\nu x(1-y)\right) \] and \[ \Om_\phi = - \frac{R\la\sqrt{2((1+\nu)^2-\la^2)}}{\Hyx} (1-x^2)y\sqrt{\nu}. \]

The form of the that metric we use here is slightly different, although entirely equivalent, to that presented elsewhere in the literature.  Relative to \cite{Pomeransky}, the $\phi$ and $\psi$ coordinates have been exchanged, to be consistent with the singly spinning solution as presented in the review \cite{ER:2006}, and the functions $F(x,y)$ and $J(x,y)$ have been replaced with $A(x,y)$ and $L(x,y)$ defined such that \[ F(x,y) = \frac{R^2 A(x,y)}{(1-\nu)^2 (x-y)^2} \eqand J(x,y) = \frac{R^2 L(x,y)}{(1-\nu)^2 (x-y)^2}.\]  The length-scale parameter $R$ is related to their $k$ by $R^2=2k^2$.

It is useful at this stage to think a little bit more carefully about the properties of the metric functions $\Axy$ and $\Lxy$.  Is it immediately apparent from the definition of $\Axy$ that we can write it in the form \begin{equation} \Axy = \Gx \ayb + \Gy \bxb \end{equation} for some $\alpha(\xi)$ and $\beta(\xi)$.  Note that there is a freedom in our choice of these functions; we can add an arbitrary multiple of $G(\xi)$ to one and subtract it from the other without affecting $A(x,y)$ itself.  It turns out that the most convenient way of doing this is to pick \[\alpha (\xi) = \nu (1-\xi^2) \left[- (1+\la^2) - \nu(1-\nu) + \la \xi (2-3\nu) - (1-\la^2)\xi^2  \right] \] and \[{\beta}(\xi) = (1+\la^2) + \la \xi (1+(1-\nu)^2) \\ - \nu\xi^2 (2\la^2+\nu(1-\nu)) - \la \nu^2\xi^3 (3-2\nu) - \nu^2\xi^4 (1-\la^2 + \nu(1-\nu)) . \] We can also do a similar thing for $\Lxy$.  If we set \[ \gamma(\xi) = \la \sqrt{\nu} (1-\xi^2) (\la - (1-\nu^2) \xi - \la\nu \xi^2) \] then we find that \[ \Lxy = \Gx \gy - \Gy \gx . \]

The ring-like coordinates can be related to two pairs of polar coordinates $(t,r_1,\phi,r_2,\psi)$ via \begin{equation} r_1 = R \frac{\sqrt{1-x^2}}{x-y} \eqand r_2 = R \frac{\sqrt{y^2-1}}{x-y},\label{eqn:planep}. \end{equation}  Note that, in these coordinates, the flat space limit takes the standard form \[ ds^2 = -dt^2 + dr_1^2 + r_1^2 d\phi^2 + dr_2^2 + r_2^2 d\psi^2 .\] The black ring has a ring-like curvature singularity at $y\rightarrow -\infty$, which is the ring $(r_1,r_2)=(0,R)$ in the polar coordinates (\ref{eqn:planep}).

\subsection{Inverse Metric} \label{sec:inv}
The inverse metric will be useful later, so we give it here for convenience, it reads
\begin{multline} \label{eqn:inv} \left( \frac{\pd}{\pd s} \right)^2 = -\frac{\Hxy}{\Hyx} \left( \frac{\pd}{\pd t} \right)^2 + \frac{(x-y)^2 (1-\nu)^2}{R^2 \Hxy} \Bigg[ \Gx \left( \frac{\pd}{\pd x} \right)^2 - \Gy \left( \frac{\pd}{\pd y} \right)^2 +\\ \frac{\Axy \left( \frac{\pd}{\pd \phi} -\Om_\phi \frac{\pd}{\pd t} \right)^2 - 2\Lxy \left( \frac{\pd}{\pd \phi} -\Om_\phi \frac{\pd}{\pd t} \right) \left( \frac{\pd}{\pd \psi} -\Om_\psi \frac{\pd}{\pd t} \right) - \Ayx \left( \frac{\pd}{\pd \psi} -\Om_\psi \frac{\pd}{\pd t} \right)^2}{(1-\nu)^2 \Gx \Gy} \Bigg].  \end{multline}  Note that \[ \frac{\Axy}{\Gx \Gy} = \frac{\ayb}{\Gy} +  \frac{\bxb}{\Gx} \] separates into $x$ and $y$ components, as do the analagous expressions for $\Ayx$ and $\Lxy$.

\subsection{Horizon}\label{sec:hor}
The metric is singular when the function $G(y)$ vanishes.  The root at \[ y = y_h \equiv \frac{-\la + \sqrt{\la^2-4\nu}}{2\nu} \] is a coordinate singularity corresponding to an event horizon.  Elvang and Rodriguez \cite{Elvang:2007bi} give a prescription for changing to new coordinates that are valid across the horizon, although it is very complicated to write the transformed metric down explicitly.  Later in this paper, we will construct an alternative set of coordinates that are valid as we cross the horizon, by looking for coordinates adapted to a particular class of null geodesics.


\subsection{Asymptotic Flatness}\label{sec:asymp}
This spacetime is (globally) asymptotically flat, but this is not manifest in the ring-like coordinates, where asymptotic infinity corresponds to the point $(x,y)=(-1,-1)$.  To see the asymptotics explicitly, we can make a change of variables $(x,y)\mapsto (\rho,\theta)$ by setting \[x=-1+\frac{2R^2}{\rho^2}\frac{1+\nu-\la}{1-\nu} \cos^2 \theta \;\;\; \mathrm{and} \;\;\; y=-1 - \frac{2R^2}{\rho^2}\frac{1+\nu-\la}{1-\nu} \sin^2 \theta,\] with $R\sqrt{(1+\nu-\la)/(1-\nu)} \leq \rho < \infty$ and $0\leq \theta \leq \pi$.  Therefore, for large values of $\rho$, the metric reduces to \[ ds^2 \approx -dt^2 + d\rho^2 + \rho^2 (d\theta^2 + \cos^2 \theta d\phi^2 + \sin^2 \theta d\psi^2),\] which is 5-dimensional Minkowski space expressed in polar coordinates, with the angular variables having the correct periodicities.  This transformation is motivated by that given in \cite{Elvang:2007bi}.

\subsection{Singly Spinning Limit}\label{sec:singly}
Since the coordinates used here vary slightly from those used in most papers on singly spinning rings (eg \cite{ER:2001}, \cite{ER:2006}, \cite{ER:2008}, \cite{Hoskisson}) it is worth showing explicitly how this reduces to the original Emparan-Reall solution.

The singly spinning limit corresponds to setting $\nu=0$.  This reduces the metric functions to the following:\[ \Gx =(1-x^2)(1+\la x), \quad H(x,y) = 1+2x\la+\la^2 \equiv H(x),\]
\[ \axb = \gx = \Lxy = 0,\quad \beta(x) = H(x) , \quad A(x,y) = H(x) G(y) \] \begin{equation} \mathrm{and} \quad \Om = \Om_\psi d\psi = -CR\frac{1+y}{H(y)}d\psi, \quad \mathrm{where} \quad C^2\equiv 2\la^2\frac{(1+\la)^3}{1-\la}.\label{eqn:Cdef} \end{equation}  The convenience of the limits here is our main motivation for working with the particular choices of $\alpha$ and $\beta$ that we made above.

The metric reduces to \begin{equation} \label{eqn:singlymet} ds^2 = -\frac{\Hy}{\Hx}(dt+\Om_\psi d\psi)^2 + \frac{R^2 \Hx}{(x-y)^2} \left[ \frac{\Gx}{\Hx} d\phi^2 + \frac{dx^2}{\Gx} - \frac{\Gy}{\Hy} d\psi^2 - \frac{dy^2}{\Gy} \right].\end{equation}

\subsection{Ergoregion}\label{sec:ergo}
For the singly-spinning black ring, the ergoregion was first described in \cite{ER:2001}.  It is straightforward to see that, in our notation, the ergosurface is where $H(y)$ vanishes, which occurs at \[y = y_e \equiv -\frac{1+\la^2}{2\la}.\]  Furthermore, we have that $ y_h < y_e < -1$, for all $\la$, so the ergoregion does indeed exist, and, like the horizon, has topology $S^1 \times S^2$ (like all surfaces $y=\mathrm{const}$ for $y \neq -1$).

Things become significantly more complicated in the doubly spinning case.  The ergosurface is the surface where $H(y,x)$ vanishes, so is a surface $y=y_e(x)$ with a more complicated shape, and we will see in the following that it sometimes also has a different topology.

Note that $H(-1,-1)=(1-\nu)(1+\nu-\la)^2 > 0$, and therefore $H(y,x) > 0$ in some neighbourhood of asymptotic infinity, so far from the ring $\pd / \pd t$ is indeed timelike as expected (as it has to be for consistency with the asymptotic analysis above!).
 
However, it can be shown that, for all $x\in[-1,1]$, $y_e(x)>y_h$, and hence the horizon is always surrounded by an ergoregion.  In the singly spinning case, $H(-1)>0$, and hence the axis $y=-1$ lies outside the ergoregion and it has ringlike topology.  However, when $\nu$ is sufficiently large, there are some values of $x$ for which $H(-1,x)<0$, and hence the ergosurface intersects the axis and can therefore no longer have the ring-like topology $S^1\times S^2$.

What is the new topology?  Note that \[H(-1,x)=H(-1,-x)=(1-\la)^2-\nu^2 + \nu x^2 \left(1-\la^2-\nu^2+2\la\nu\right)\] is even as a function of $x$, and that therefore \[ H(-1,1)=H(-1,-1) = (1-\nu)(1+\nu-\la)^2 > 0. \]  Thus, for all allowed values of $\la$ and $\nu$ we have that the point at the centre of the ring lies outside of the ergoregion.  As $\nu \rightarrow 1$ (and hence $\la\rightarrow 2$), the size of the ergoregion becomes larger and larger, but there is always a region near to the centre of the ring that remains outside it.  Thus, the ergosurface topology is that of two disconnected 3-spheres, $S^3\cup S^3$.

\begin{figure}[htb]
\begin{center}
\begin{minipage}[c]{0.9\textwidth}
\includegraphics[width=\textwidth]{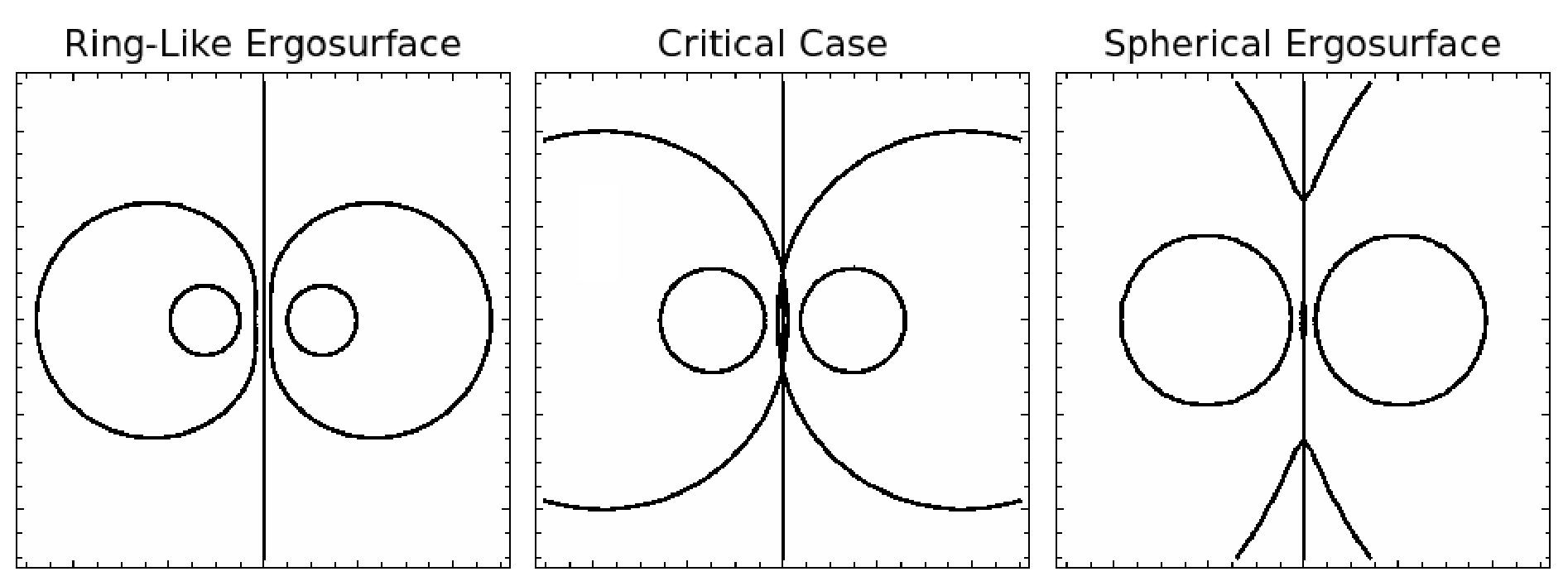}
\caption{ \label{fig:ergoall}\footnotesize{Two-dimensional projection of the shape of the ergoregion in the case $\nu=1/9$, for $\la = 7/9$ ($S^1\times S^2$ ergosurface), $\la = 8/9$ (critical case) and $\la = 1$ ($S^3 \cup S^3$ ergosurface).  The inner circles are the edge of the horizon, the outer lines the ergosurface and the central line the axis $y=-1$. (Plotted in $r_1$, $r_2$ coordinates.)}}
\end{minipage}
\end{center}
\end{figure}

Note that $H(-1,x)$ is minimum at $x=0$, so to determine where the change of topology occurs we need to look at the case where \[ H(-1,0)= 1+\la^2 - \nu^2 - 2\la = 0.\] This occurs when $\la=1-\nu$.  Note that we must have $\nu \leq 3-2\sqrt{2} \simeq 0.17$ for it to be possible to have this condition satisfied.  Here, we have that \[ H(-1,x) = 4\nu^2 x^2 (1-\nu), \] so the ergosurface touches the $y=-1$ axis on the circle $x=0$, $y=-1$.  In the plane polar type coordinates (\ref{eqn:planep}), the locus of points where the ergosurface pinches is at $r_1 = R$, $r_2 = 0$, which makes clear that this is indeed a circle.  We will see later (\S \ref{sec:axisgeo}) that there exist null geodesics that orbit around this circle.  Figure \ref{fig:ergoall} shows a 2D projection of the shape of the ergoregion in this case.
\begin{figure}[htb!]\begin{center} \leavevmode 
\input{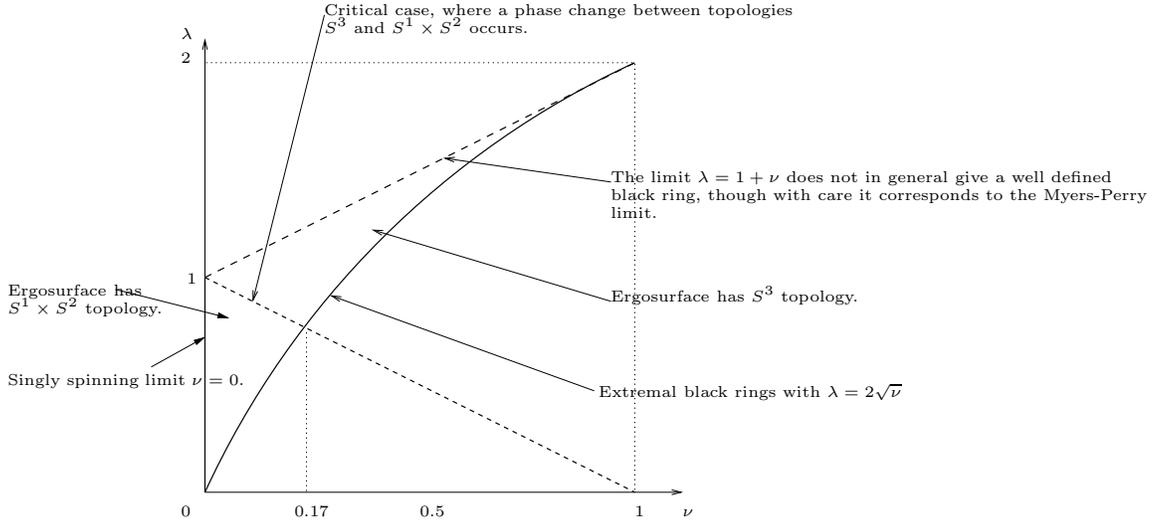}
\caption{ \label{fig:ergophase} \small{The parameter space for doubly spinning black rings.} }
\end{center}
\end{figure}

Finally, there is a nice intuitive way to think about why the ergoregion takes this form.  We can think, rather loosely, of the black ring as a Kerr black hole at each point around the $S^1$.  When the Kerr black hole is rotating rapidly (corresponding to rapid $S^2$ rotation of the black ring), its ergoregion becomes increasingly elliptical, so that eventually an observer near the centre of the ring feels frame dragging from the $S^2$ rotations on opposite sides of him simultaneously.  The effects cancel near the centre of the ring, leaving a region which does not lie in the ergoregion.  To summarise, Figure \ref{fig:ergophase} shows the parameter space for all allowed doubly-spinning black rings.
\newpage
\section{Geodesic Structure} \label{sec:geo}
Hoskisson \cite{Hoskisson} has studied in detail certain classes of geodesics for the singly spinning black ring.  In particular, he studies analytically families of geodesics restricted to the axes $y=-1$ and $x=\pm 1$, as well as performing numerical investigations into some more general possibilities.  In this paper we mainly concentrate on a different class of geodesics, which we can also find explicitly.  We show that, in the full doubly spinning case, the Hamilton-Jacobi equation is separable for null, zero energy geodesics.  This condition can only be satisfied in the ergoregion of the spacetime.

We go on to discuss the behaviour of geodesics that result from this.  Finally, we also give a brief discussion of the possibilities for analysing axis geodesics in this case.  The two axes are fixed-point sets of the isometries generated by the respective axial Killing vectors, and hence are totally-geodesic submanifolds of the geometry.  Thus, as in the singly spinning case, the geodesic equations can be reduced to ODEs on these surfaces.

\subsection{Conjugate momenta}
As usual, we look for geodesics by noting that they are extremal curves of the Lagrangian \[\lag = \frac{1}{2}g_{ab}\dot{x}^a \dot{x}^b ,\] where a dot denotes differentiation with respect to an affine parameter $\tau$.  The conjugate momenta for this Lagrangian are \begin{eqnarray}
E \equiv - p_t &=& \frac{\Hyx}{\Hxy}(\dot{t}+\dot{\Om})\label{eqn:mom}\\
\Phi \equiv p_\phi &=& -\Om_\phi E - \frac{R^2}{\Hyx (x-y)^2 (1-\nu)^2}(-\Ayx \dot{\phi}+\Lxy \dot{\psi}) \nonumber\\
\Psi \equiv p_\psi &=& -\Om_\psi E - \frac{R^2}{\Hyx (x-y)^2 (1-\nu)^2}(\Axy \dot{\psi}+\Lxy \dot{\phi}) \nonumber\\
p_x &=& \frac{R^2 \Hxy \dot{x}}{(x-y)^2 (1-\nu)^2 \Gx} \nonumber\\
p_y &=& \frac{-R^2 \Hxy \dot{y}}{(x-y)^2 (1-\nu)^2 \Gy} \nonumber
\end{eqnarray} where $\dot{\Om} \equiv \Om_\psi \dot{\psi} + \Om_\phi \dot{\phi}$.  The vector fields $\pd / \pd t$, $\pd / \pd \phi$ and $\pd / \pd \psi$ are Killing, so the conjugate momenta $-E$, $\Phi$ and $\Psi$ associated with them are conserved along any geodesics.

\subsection{The Hamilton-Jacobi Equation} \label{sec:hj}
The Hamilton-Jacobi approach can offer us some interesting insights into the geodesic structure of a system.  Let $\ham (x^a,p_b)$ be the Hamiltonian for particle motion in this background, derived from the Lagrangian $\mathcal{L}(x^a,\dot{x}^b)$ in the usual way through a Legendre transformation
\[ \ham (x^a,p_b) \equiv p_a \dot{x}^a - \mathcal{L}(x^a,\dot{x}^b) = \frac{1}{2} g^{ab}p_a p_b .\]

The aim of this approach is to give us an additional constant of motion.  The system is 5-dimensional, so we need 5 constants of motion in order to be able to completely integrate it.  Applying Noether's theorem to the Killing vectors $\pd / \pd t$, $\pd / \pd \psi$ and $\pd / \pd \phi$ has already given 3 of them, and we also impose the mass shell condition $g^{ab} p_a p_b = -\mu^2$ which gives a fourth.  Therefore, one more is required.

We look for additively separable solutions of the HJ equation \begin{equation} \label{eqn:hj} \frac{\pd S}{\pd \tau} + \mathcal{H}\left(x^a,\frac{\pd S}{\pd x^b}\right) = 0. \end{equation} Given our prior knowledge of 4 constants of motion, we make an ansatz \[ S(\tau,t,x,y,\psi,\phi) = \frac{1}{2}\mu^2 \tau - Et + \Phi \phi + \Psi \psi + S_x(x) + S_y(y),\] where $\tau$ is an affine parameter along a geodesic, and $S_x$, $S_y$ are arbitrary functions of $x$ and $y$ respectively.  We can choose $\tau$ to be proper time in the timelike geodesic case.  We hope that this ansatz will leave the HJ equation (\ref{eqn:hj}) in a separable form.

Inserting this ansatz into (\ref{eqn:hj}) gives, after some rearrangement, \begin{multline} \Gx\dSdx-\Gy\dSdy = \\\frac{R^2 \Hxy}{(1-\nu)^2(x-y)^2} \left(- \mu^2 + \frac{\Hxy}{\Hyx}E^2\right)  -\frac{\Hxy \Hyx}{\Axy \Ayx + \Lxy^2}\big[\Axy(\Phi+\Om_\phi E)^2 -  \\ 2\Lxy (\Phi+\Om_\phi E)(\Psi+\Om_\psi E)  - \Ayx(\Psi+\Om_\psi E)^2\big] .\label{eqn:hj1}\end{multline} At first glance, it appears that there is little hope of separating this.  However, it is possible to make some progress, using relations between the metric functions that are not immediately apparent from the solution as presented in \cite{Pomeransky}:\begin{itemize}
\item Firstly, note the identity \begin{equation} \Axy \Ayx + \Lxy^2 \equiv \Gx \Gy \Hxy \Hyx (1-\nu)^2 .  \label{eqn:id}\end{equation} This simplifies (\ref{eqn:hj1}) to \begin{multline}\Gx\dSdx-\Gy\dSdy= \frac{R^2 \Hxy}{(1-\nu)^2(x-y)^2}\left(-\mu^2 + \frac{\Hxy}{\Hyx} E^2\right) \\
- \frac{\left[\Axy(\Phi+\Om_\phi E)^2 - 2\Lxy (\Phi+\Om_\phi E)(\Psi+\Om_\psi E)  - \Ayx(\Psi+\Om_\psi E)^2\right]}{\Gx\Gy(1-\nu)^2} \label{eqn:hj2}.
\end{multline}
\item Writing \begin{equation} \Axy = \Gx \ayb + \Gy \bxb \end{equation} allows us to separate the $\Phi^2$ and $\Psi^2$ terms of (\ref{eqn:hj2}).
\item It is also possible to separate the $\Phi\Psi$ term using the relation \begin{equation} \Lxy = \Gx \gy-\Gx\gx. \end{equation}
\item It is not possible, in general, to separate the $\mu^2$ and $E^2$ terms, nor those involving $\Om_\phi$ and $\Om_\psi$.\end{itemize}

Therefore, the only separable solutions in these coordinates correspond to null ($\mu=0$), zero energy ($E=0$) geodesics, with $S_x$ and $S_y$ satisfying \begin{equation}\Gx\dSdx- \frac{-\bxb \Phi^2 - 2\gx \Phi\Psi + \axb \Psi^2}{(1-\nu)^2 \Gx} = \Gy\dSdy - \frac{\ayb\Phi^2 - 2\gy \Phi\Psi - \byb \Psi^2}{(1-\nu)^2 \Gy} . \label{eqn:sep}\end{equation}  Given this separation of variables, we can then immediately write \[ \mathrm{LHS} = \mathrm{RHS} = \frac{c}{(1-\nu)^2} \] for some constant $c$.  This describes all possible null, zero energy geodesics.  $c$ is the extra constant required to allow the geodesic equations to be completely integrated in this case, unlike the Noether constants associated with Killing vectors it is quadratic in the momenta (see Section \ref{sec:sym}).  These geodesics are physically realisable in the ergoregion, where $\pd / \pd t$ is spacelike.

It is worth noting that the separability of the HJ equation is a coordinate dependent phenomenon.  In particular, the HJ equation describing flat space geodesics is not separable in ring-like coordinates.  In fact, the general solution for flat space geodesics can be written in ring-like coordinates as \[ S(t,x,y,\phi,\psi;\tau) = \frac{1}{2}\mu^2 \tau - Et + \frac{R}{x-y} \left[ R_1\sqrt{1-x^2} \cos(\phi-\phi_0) + R_2 \sqrt{y^2-1} \cos(\psi-\psi_0)\right] + \mathrm{const}\] with $\phi_0$, $\psi_0$, $R_1$, $R_2$, $\mu^2$ and $E$ arbitrary constants.

\subsection{Analysis of Paths of Ergoregion Geodesics}\label{sec:ergogeo}
Given the results of Section \ref{sec:hj}, we can study the paths of zero energy, null geodesics in the ergoregion explicitly.  Since the zero energy, null condition is only realisable in the ergoregion, an observer moving along such a geodesic cannot move outside of the ergoregion, though could potentially fall in through the horizon.

The separated Hamilton-Jacobi equation gives us that \begin{equation}\label{eqn:ergox} \frac{R^4 \Hxy^2}{(x-y)^4 (1-\nu)^2} \dot{x}^2 + U(x) = 0 \end{equation} and \begin{equation} \label{eqn:ergoy} \frac{R^4 \Hxy^2}{(x-y)^4 (1-\nu)^2} \dot{y}^2 + V(y) = 0 \end{equation} where \[ U(x) = - \left[-\bxb \Phi^2 - 2\gx \Phi\Psi + \axb \Psi^2 + c \Gx \right] \] and \[ V(y) = - \left[\ayb\Phi^2 - 2\gy \Phi\Psi - \byb \Psi^2 + c \Gy \right] . \] These equations give coupled effective potential formulations for the motion, and we can use them to deduce the behaviour of this class of geodesics.  When dealing with effective potentials, it it usually useful to rearrange the equation such that one of the Noether constants (usually the energy) sits alone on the RHS, making it easy to understand how things change as that parameter varies.  Unfortunately, this is not possible in all cases here.

Note that, at least implicitly, we can use these equations to find $x$ as a function of $y$.  Dividing through, and noting that the prefactors with mixed $x$ and $y$ dependence cancel, we have that \begin{equation}\label{eqn:dxdy} \left(\frac{dx}{dy}\right)^2 = \frac{U(x)}{V(y)} \quad \Rightarrow \quad \int^x \frac{dx}{\sqrt{-U(x)}} = \int^y \frac{dy}{\sqrt{-V(y)}},\end{equation} which gives us what we need.

Although these two effective potential equations are coupled to each other, the coupling arises only through the strictly positive pre-factor of the kinetic term.  Thus, the coupling has no effect on whether the potential is attractive or repulsive, or on its turning points.  Therefore, we can effectively treat the two parts independently when studying the qualitative behaviour of geodesics.

\subsubsection{Singly spinning case}\label{sec:singlygeo}
To begin with, it is easier to study these ergoregion geodesics in the singly spinning case $\nu=0$.  Here, the equations (\ref{eqn:ergox}) and (\ref{eqn:ergoy}) reduce to \begin{equation}\label{eqn:ergoxsing} \dot{x}^2 + \frac{(x-y)^4 }{R^4 \Hx^2} \left[ \Phi^2\Hx-c\Gx \right] = 0 \eqand \dot{y}^2 + \frac{(x-y)^4}{R^4 \Hx^2} \left[\Psi^2 \Hy - c\Gy \right] =0. \end{equation} Note that the ergoregion is the region $ -\frac{1}{\la} < y < - \frac{1+\la^2}{2\la} $ here, with topology $S^1 \times S^2$.  The $y$ motion is of the most immediate interest, since that governs how close to the horizon the path lies.

As noted by Hoskisson \cite{Hoskisson}, care is needed when we get near to the axes $y=-1$ or $x=\pm 1$, since the angular coordinates $\psi$ or $\phi$ respectively become singular there.  However, this is a coordinate singularity, originating from the singularity at the origin in the plane polar coordinates (\ref{eqn:planep}), and hence we expect that taking limits like $y \rightarrow -1$ should be valid.  This can be confirmed in a straightforward (though messy) manner using the transformations to cartesian coordinates described in \cite{ER:2006}.

There are several cases to consider:
\paragraph{Case $c=0$:} Recall that $c$ is the separation constant from the Hamilton-Jacobi equation, so it parametrises a set of geodesic curves.  Now, we must have $c \geq 0$ to have an effective potential for $x$ that is non-positive somewhere, and hence some allowed solutions, so it is natural to begin with the bounding case $c=0$ and see what that gives us.  Note that $\Hx>0$ for all $x\in [-1,1]$, so in this case we also require $\Phi=0$ for any solution.  We must then have $\Psi\neq 0$ to get a path at all, and thus are left with the effective potential formulation \[\dot{x}^2 = 0 \eqand \dot{y}^2 + \frac{(x-y)^4 \Psi^2\Hy}{R^4 \Hx^2} = 0.\]  We have $\Hy <0$ everywhere inside the ergoregion, and $\Hy=0$ on the ergosurface, so the only turning point $\dot{y} = 0$ of the geodesic lies on the ergosurface.  The other coordinate $x$ is constant along these geodesics, so acts as an arbitrary constant rather than a dynamical variable in the $y$ equation, and in fact has no qualitative effects on the paths.  These solutions must correspond to geodesics that have come out of the white hole horizon in the past, move outwards away from the black ring until they just touch the ergosurface and then turn round and fall back into the black hole horizon in finite parameter time in the future.

\begin{figure}[hbt]
\begin{center}
\begin{minipage}[c]{0.9\textwidth}
\includegraphics[width=0.49\textwidth]{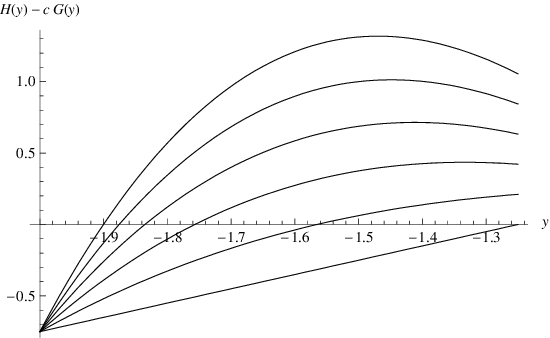}
\includegraphics[width=0.49\textwidth]{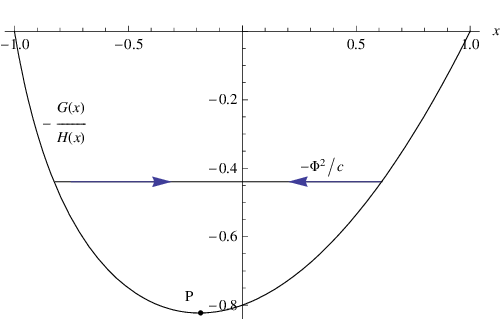}
\caption{\footnotesize{\label{fig:singlypot}\label{fig:GoverH}\textbf{(a)} $\Hy - \bar{c} \Gy$ plotted against $y$ in the ergoregion ($-2 \leq y \leq -\frac{5}{4}$) for $\la=\tfrac{1}{2}$, $\nu=0$, for $\bar{c}=0,1,2,3,4,5$.  The potential in each case is bounded below by the $\bar{c}=0$ potential (the bottom line). \textbf{(b)} The $x$-motion effective potential $-G(x)/H(x)$ plotted against $x$.  This potential determines the allowed values of the constant $-\Phi^2/c$, an example path is plotted. (Figure has $\la=\frac{1}{2}$, $\nu=0$.)}} 
\end{minipage}
\end{center}
\end{figure}

\paragraph{Case $c>0$ and $\Phi=0$:} Here it is less easy to be explicit, but we can deduce the behaviour of these geodesics by relating them to the $c=0$ case.  The relevant equations are \[ \frac{R^4 \Hx^2}{(x-y)^4} \dot{x}^2 -  c\Gx = 0 \eqand \frac{R^4 \Hx^2}{(x-y)^4 \Psi^2 } \dot{y}^2 + \left[\Hy - \bar{c} \Gy \right] =0,\] where $\bar{c}\equiv c/\Psi^2$.  Since $\Gy<0$ outside the horizon, the effective potential for $y$-motion in the $c>0$ case is bounded below by that in the $c=0$ case, with equality only at $y=-1$ and $y=-\frac{1}{\la},$ that is at the horizon.  Thus, the geodesics in this case have the same qualitative behaviour, but stop short of the ergosurface before falling inwards again.  Figure \ref{fig:singlypot}(a) shows how the turning point of the geodesic (occurring where $\Hy - \bar{c}\Gy=0$) moves inwards as $c$ is increased.

Note that in this case, $x$ also varies, which makes integrating the motion explicitly far more difficult, though it has no real effect on the qualitative  form of the motion in $y$.  Since $c\Gx \geq 0 $ everywhere,  $x$ can take any value in $[-1,1]$.  This corresponds to the particle continually rotating around the $S^2$ part of the horizon as it moves in $y$.

\paragraph{Case $c>0$ and $\Phi>0$:}
In the singly spinning case, $\Phi$ does not enter into the effective potential for $y$, and therefore does not change the turning points in the $y$ motion.  However, the $x$ dynamics are now more interesting.  We can write the effective potential equation for $x$ as \[ \frac{R^4 \Hx}{c (x-y)^4} \dot{x}^2 - \frac{\Gx}{\Hx} = -\Phi^2/c,\] and hence see that there is a restriction on the values of $x$ that are possible.  For $\Phi^2/c=0$, any values of $x$ are allowed, but as $\Phi^2/c$ is increased, $x$ is restricted to an increasingly narrow range of values, corresponding to a centrifugal repulsion keeping the particle away from the axis $x=\pm 1$.  Rather than continuously rotating around the $S^2$, the particle follows a more complicated path, bouncing back and forth between two different extremal values of $x$.  This also gives us an upper bound on the values of $\Phi^2/c$ that are allowed, as shown by Figure \ref{fig:GoverH}(b).
There is a non-trivial fixed point in the $x$ potential (marked $P$ in Figure \ref{fig:GoverH}(b)), corresponding to an orbit at fixed $x$ when $\Phi^2/c$ takes its maximum allowed value.  It is messy to solve the cubic required to compute the exact location of the fixed point, and the corresponding maximum value of $\Phi^2/c$, and we do not do it here.

\subsubsection{Doubly spinning} \label{sec:dsgeo}
This concludes the possibilities for the singly spinning ring, and describes all of the possibilities for the behaviour of zero energy, null geodesics lying inside the ergosurface.  We will now attempt to move beyond this, and study the same thing in the doubly spinning version.  Unfortunately, it is less easy to be explicit here, so we will limit ourselves to showing the existence of the geodesics, and discussing their properties in a couple of special cases.  The relevant effective potential equations are (\ref{eqn:ergox}) and (\ref{eqn:ergoy}).

In the previous section, we showed explicitly that the geodesics turned around before reaching the ergosurface (or in the limiting case, on the surface itself).  However, it is not strictly necessary to do this, since it can be deduced from well-known properties of geodesics.  Having found a section of a null, zero energy geodesic, we know that we can extend the geodesic indefinitely both forwards and backwards in time in a unique way, unless it hits a singularity (indeed, this is how one usually defines a singularity in a spacetime).  Furthermore, the geodesic extension of this curve must remain a null, zero energy geodesic.  Since the zero energy, null condition cannot be satisfied outside of the ergoregion, a particle travelling along such a geodesic cannot possibly pass through the ergosurface, and can only leave the ergoregion by passing through a horizon.

Now let's move on to consider some particular cases:
\paragraph{Case $\Phi=0$}
The full equations simplify significantly if we set one of the angular momenta to zero, specifically $\Phi$ (recall from the singly spinning case that there were no allowed zero-energy paths with $\Psi=0$; it is straightforward to show that the same applies here).  This leaves us with \[ U(x) = -\axb \Psi^2 - c\Gx \;\; \mathrm{and} \;\; V(y) = \byb \Psi^2 - c\Gy, \] essentially leaving us with one tunable parameter $\bar{c} \equiv c / \Psi^2$.

\begin{figure}[htb]
\begin{center}
\begin{minipage}[c]{0.9\textwidth}
\includegraphics[width=0.49\textwidth]{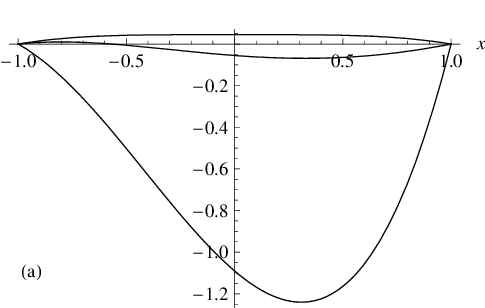}
\includegraphics[width=0.49\textwidth]{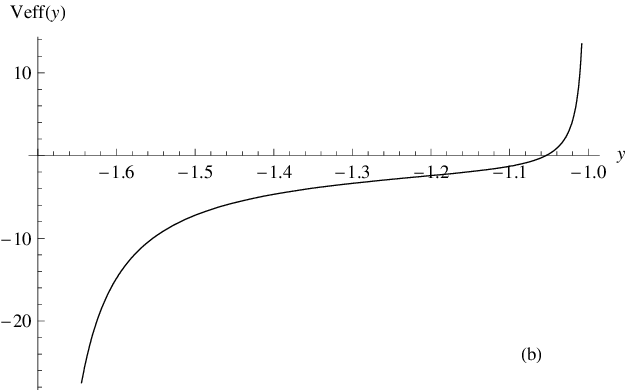}
\caption{\footnotesize{\label{fig:doublyypot}\label{fig:doublyxpot}\textbf{(a)} Possible behaviours of the effective potential $U(x)$ for the doubly spinning ring in the case $\Phi=0$, for 3 different values of $\bar{c}=0,\frac{1}{10},\frac{276}{243}$.  The top curve gives no allowed geodesics, the bottom one allows all values of $x$. \textbf{(b)} The effective potential $\Veff(y) = -{\beta}(y)/G(y)$ for $y$-motion.  The horizon is located at the vertical axis on the left.  Both parts of this figure are plotted for $\la=\frac{1}{9}$, $\nu=\frac{7}{9}$, but the shape of the potentials is insensitive to changes in $\la$, $\nu$.}}
\end{minipage}
\end{center}
\end{figure}

Firstly, let us consider the motion in $x$.  Qualitatively there are 3 different possibilities for the potential $U(x)$ in this case, as shown in Figure \ref{fig:doublyxpot}(a).  The cases are: \begin{description}
\item[No values of $x$ allowed.]  If we have $U(x) >0$ for all $x$, then there can be no geodesics.  This occurs iff $U'(1) <0$, or equivalently \[ \bar{c} < \frac{\nu}{1+\la+\nu} \left[ 2(1-\la)+\nu(1-\nu)+3\la\nu \right].\]  This fixes a lower bound for $\bar{c}$.
\item[Some values of $x$ allowed.] If $U'(1) > 0$, but $U'(-1)>0$ as well, there are allowed geodesics, but they are restricted to a certain range in $x$, with the very `outside' of the ring excluded.  This occurs if \begin{equation} \frac{\nu}{1+\la+\nu} \left[ 2(1-\la)+\nu(1-\nu)+3\la\nu \right] <  \bar{c} < \frac{\nu}{1+\nu-\la} \left[ 2(1+\la) - 3\la\nu + \nu(1-\nu) \right].\label{ine:somex}\end{equation}
\item[All values of $x$ allowed.] The $x$-range of the geodesics is entirely unrestricted, and they are free to loop all of the way around the $S^2$ of the ring.  This occurs if \[ \bar{c} \geq \frac{\nu}{1+\nu-\la} \left[ 2(1+\la) - 3\la\nu + \nu(1-\nu) \right].\] \end{description}

Note that the middle case (\ref{ine:somex}) does not occur for the singly spinning ring, and the analysis above reduces to demanding that $c\geq 0$.
For the $y$-motion, it turns out that the qualitative form of the motion is exactly the same as in the singly-spinning case.  Note that \[V(y_h) = {\beta}(x) \Psi^2 < 0 ,\] so the potential is negative in some neighbourhood of the origin, and there is nothing (locally) to block a geodesic from crossing it.   Given this, the easiest way to study the behaviour away from the horizon is to express the potential equation as \[ \frac{R^4 \Hxy^2 \Psi^2}{(x-y)^4 (1-\nu)^2 (-\Gy)} \dot{y}^2 + V_{\mathrm{eff}}(y) = -\bar{c} \] where $ \Veff(y) = -\byb/\Gy$.  To analyse the system, we need to study $\Veff(y)$ in the ergoregion.  Finding roots explicitly is hard, since it requires finding roots of a complicated quartic equation, but it can be shown (by differentiating and using the bounds on allowed values of $\la$, $\nu$ in various ways) that outside the horizon, for all values of $\la$ and $\nu$, $\Veff'(y)  > 0$ and hence there are no fixed points of the potential.  Therefore there can be no closed orbits.  As described above, we know from general principles of geodesics that all of these geodesics must turn around before getting outside of the ergoregion, so we know that $\Veff(y)$ must vanish for some $y<y_e(x)$.  However, this is only true for for a certain subset of $x$ values, and thus, there is a restriction on the allowed $x$ values near to the turning point of the geodesic.  We know that this must be consistent with the restrictions on $x$ obtained from analysing the $x$-potential.

\paragraph{General $\Phi$}
Note that $U(\pm 1) = (1-\nu)^2 (1+\nu\pm \la)^2 \Phi^2$, which is strictly positive for $|\Phi| >0$.  Therefore, the $x$ potential can no longer be categorised by finding derivatives at either end of the allowed range of $x$ values.  Instead, it is necessary to find turning points of the quartic $U(x)$ explicitly in order to find the range of $x$ values where $U(x)\leq 0$.  This is extremely messy, so we will not do it here.  However, there is a clear qualitative difference here; as soon as $|\Phi| >0$ there is a centrifugal barrier preventing these geodesics from touching the plane $x=\pm 1$.  Otherwise, the basic qualitative result is the same as in the singly spinning case; there is an upper bound on the allowed value of $\Phi^2/c$ in order to get allowed orbits of any kind.

The $y$ motion here is more complicated still, however numerical investigations suggest that no new behaviour occurs; that is all geodesics come out of the white hole and fall back into the black hole in finite proper time.  We do not present a detailed analysis of this case here.

\subsection{Other analytically tractable geodesics}\label{sec:axisgeo}
While it is extremely unlikely to be possible to study all geodesics of this metric analytically, some progress can be made with finding geodesics that have particular symmetry.  In particular, it is possible to find geodesics lying entirely within surfaces that are fixed-point sets of the axial Killing vectors $\pd / \pd\phi$ and $\pd / \pd\psi$.  These surfaces are totally geodesic submanifolds, in that any geodesic that lies tangent to the submanifold at some point must lie entirely within the submanifold.  Typically, this introduces an extra constraint on the equations of motion, and reduces the problem to solving an ODE, the qualitative behaviour of which can be analysed via effective potential techniques.

Since these geodesics are not the main focus of this paper, we have not performed a detailed analysis of the possible different kinds of behaviour, but rather derived the appropriate effective potential equations for the motion, as well as commenting on some interesting generalities and special cases.  A full classification of all possibilities would be extremely complicated, since there is a large parameter space (any of $E$,$\la$,$\nu$,$\mu$ and one of $\Phi$ and $\Psi$ can vary), and the complexity of the potentials means that numerical graph plotting is the only reasonable approach to finding the shape of potentials in most cases.

\subsubsection{Geodesics on the $\psi$ axis}
The axis of $\psi$ rotation is the surface $y=-1$.  On this surface, geodesics can be described by a restricted version of the Hamilton-Jacobi equation, an ODE in $x$ only.  We must set $\Psi=0$ for consistency, since this is the fixed point set of $\pd/\pd \psi$.

By taking the limit $y\rightarrow -1$, and setting $p_y=0$ in the Hamilton-Jacobi equation (\ref{eqn:hj2}), we obtain an effective potential formulation \[ \frac{1}{2} \dot{x}^2 + \Veff (x) = 0\] where 
\begin{multline} \label{eqn:xpot} \Veff (x) = \frac{ (1+x)^2 (1-\nu)^2}{2R^2} \Bigg[ \frac{(1+x)^2 (\Phi + \Om_\phi E)^2}{ R^2 H(x,-1)^2} \left(\beta (x) - \frac{\nu (2+ \nu(1-\nu) + \la (2-3\nu)) G(x) }{1-\la+\nu} \right) \\ + \frac{\mu^2 G(x) }{H(x,-1)} - \frac{E^2 G(x) }{H(-1,x)} \Bigg]. \end{multline}  Care is needed in taking this limit, since the coordinates are singular there, but by transforming the coordinates $(y,\psi)$ into a set of cartesian coordinates, it can be shown that the limit is well defined.

Note that, from (\ref{eqn:xpot}) we have \[ \Veff (1) = \frac{8 (1-\nu)^2 \Phi^2}{ R^4 (1+\la+\nu)^2} , \] and hence, for $|\Phi|>0$ no geodesic along this axis can ever reach the origin $x=1$.  This is physically obvious; a particle with a conserved angular momentum about some axis cannot possibly hit that axis or it would not have any angular momentum.

\paragraph{Singly Spinning}
In the singly spinning case, it is convenient to re-write this as \begin{equation} \frac{R^2 (1-\la)^2}{ G(x) (1+x)^2} \dot{x}^2 + \frac{(1-\la)^2}{H(x)} \left[ \mu^2 + \frac{(1+x)^2 \Phi^2}{R^2 G(x)} \right] = E^2 , \label{eqn:ssaltpot}\end{equation} though this breaks down at $x=\pm 1$.

This case is described in detail by Hoskisson \cite{Hoskisson}.  Note that, up to notation, our equations, obtained by taking limits of the doubly spinning versions, are indeed the same as (36) and (37) in \cite{Hoskisson}.  The latter equations were obtained directly from the singly spinning metric, rather than as a limit of a doubly-spinning effective potential, so this is a useful consistency check.

\paragraph{Doubly Spinning}
In the doubly spinning case, there is no convenient equivalent of (\ref{eqn:ssaltpot}), since $H(-1,x)$ can change sign, depending on the sign of $1-\nu-\la$.  Thus it is easiest to work with (\ref{eqn:xpot}) directly.

Expanding around asymptotic infinity $x=-1$ gives \[ \Veff(x) \sim -\frac{(E^2-\mu^2)(1-\nu)}{1+\nu-\la} (1+x)^3 + O((1+x)^4) < 0 \] and therefore geodesics can propagate in from infinity with any values of $\Phi$ and $E>\mu$, as expected for any asymptotically flat spacetime.

In the null case, it turns out that for $|\Phi|>0$, $\Veff(x)$ always has precisely one root in $(-1,1)$ (there are 3 parameters $\la$, $\nu$, $\Phi/E$ to consider).  Thus, the only possible type of geodesic motion is a geodesic propagating in from infinity, and turning round at some $x<1$; again this can be interpreted as a centrifugal barrier at the axis of $\phi$ rotation.  The geodesic still passes through the middle of the ring in this case, but misses the origin (this feels impossible from the point of view of 3 spatial dimensions, but the extra dimension allows it to happen).

In the timelike case, things are more complicated, since there are now four free parameters $\la$, $\nu$, $E$, $\Phi$ to consider.  We do not present this case in detail here.

\begin{figure}[htb]\begin{center}\begin{minipage}[c]{0.5\textwidth}
 \includegraphics[width=\textwidth]{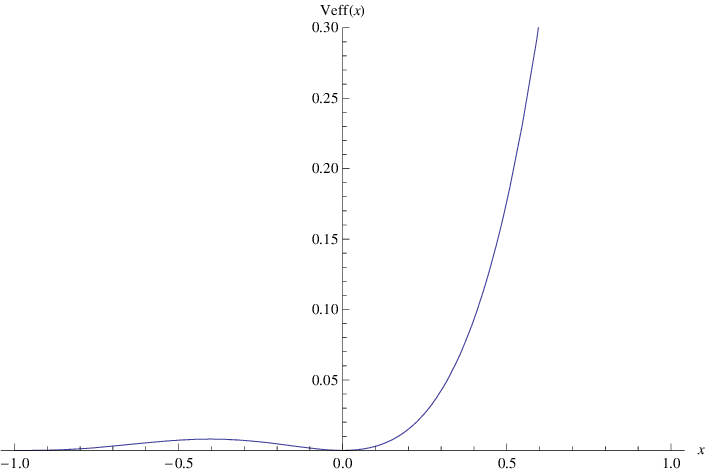}
\caption{\footnotesize{\label{fig:critpot}The effective potential $\Veff(x)$ for zero energy, null geodesics along the axis in the critical case.  We see that the only possible orbit is a stable circular one at $x=0$. (Plot has $\nu=1/9$, $\la=8/9$, $\Phi=1$)}}\end{minipage}\end{center}\end{figure}

\paragraph{Critical Case $\la = 1-\nu$}
Some particularly interesting behaviour occurs in the critical case $\la = 1-\nu$, where the ergoregion `pinches'.  Here, the potential in the case $\mu=0=E$ is given by \[ \Veff (x) = \frac{x^2 (1+x)^4 (1-\nu)^2 \Phi^2}{ 4R^4 H(x,-1) }, \] which means that there is a minimum at $x=0$, and hence a stable particle orbit there (see Figure \ref{fig:critpot}).  Thus, in this very special case, a lightlike particle can follow a circular orbit at $(r_1,r_2)=(0,R)$, on the edge of the ergoregion.

\subsubsection{Geodesics on the $\phi$ axis}
The axis of $\phi$ rotation is the surface $x=\pm 1$, which can be thought of as the equatorial plane of the ring.  In an analagous way to the previous section, we can describe geodesics on this surface by an ODE in $y$, setting $\Phi=0$ for consistency.  The two cases $x=1$ and $x=-1$ must be treated separately, they correspond to geodesics inside or outside the ring.  We can write \[ \frac{1}{2} \dot{y}^2 + \Veff^\pm (y) = 0 \] where \begin{multline} \label{eqn:ypotpm} \Veff^\pm (y) = \frac{(1\mp y)^2 (1-\nu)^2}{2R^2} \Bigg[ -\frac{\mu^2 G(y) }{H(\pm 1,y)} + \frac{E^2 G(y)}{H(y,\pm 1)} + \\ \frac{(1\mp y)^2 (\Psi + \Omy E)^2}{R^2 H(\pm 1,y)^2} \left( {\beta}(y) -\frac{\nu (2+ \nu(1-\nu) \mp \la (2-3\nu)) G(y) }{1\pm \la+\nu} \right) \Bigg] \end{multline} are the potentials in the cases $x=\pm 1$.

In both cases, we have \[ \Veff^\pm (y_h) = \frac{(1\mp y_h)^4 (1-\nu)^2 (\Psi + \Omy(\pm1,y_h) E)^2}{2R^4 H(\pm,y_h)^2} {\beta}(y_h) < 0 , \] since ${\beta} (y_h) <0$ for all $\la$, $\nu$.  Thus, there is locally nothing to prevent a geodesic crossing the horizon.  Note also that \[ \Veff^+ (-1) =\frac{8 (1-\nu)^2 \Psi^2}{ R^4 (1+\nu-\la)^2} \] which is positive for $|\Psi|>0$, and thus, as expected, no geodesics with non-zero $\Psi$ can pass through the origin.

More detailed analysis of the various cases is required to determine the paths that are possible.  Hoskisson \cite{Hoskisson} has analysed this in detail in the singly spinning case, and it seems probable that no qualitatively new behaviour occurs here, since $\Om_\phi$, the `second' rotation, vanishes on $x=\pm 1$ in any case.

\section{New Coordinate Systems} \label{sec:coord}
Until recently, there had been no attempt in the literature to construct the maximal analytic extension of a spacetime containing a black ring.  This is perhaps partly because such a result has little obvious application in attempts to quantise gravity, for example entropy calculations are not affected by what might or might not lie behind the horizon.  However, it is interesting to see how the possibilities for extension compare to that in other known systems.  Recently, Chrusciel and Cortier \cite{Chrusciel:2008} have derived the unique maximal analytic extension of the singly spinning ring, and found that its structure is similar to the Kruskal maximal extension of the Schwarzchild spacetime.  One might postulate from this that the spacetime in the doubly spinning case might have an extension similar to that of Kerr.

The first step in extending the spacetime is to construct a set of coordinates that cover the future black hole horizon.  This has been done for the singly spinning ring by Emparan and Reall \cite{ER:2001}, \cite{ER:2006}, and for the doubly spinning ring by Elvang and Rodriguez \cite{Elvang:2007bi}.  The coordinates $(\bar{t},x,\bar{\phi},y,\bar{\psi})$ of \cite{Elvang:2007bi} are defined by setting
\begin{equation}
d\bar{\phi} = d\phi - \frac{A}{y-y_h} dy, \quad d\bar{\psi} = d\psi - \frac{B}{y-y_h} dy \eqand d\bar{t} = dt - \frac{C}{y-y_h} dy .\label{eqn:elvrodcoords},
\end{equation}
and attempting to find real constants $A$, $B$, $C$ such that divergences at the horizon in metric components cancel.  This works (with an additional quadratic term needed in the extremal case $\la=2\sqrt{\nu}$), and therefore proves that the horizon is regular, but makes it hard to write down the transformed metric in a form that is manifestly regular at the horizon, to the extent that this has not been done in the literature.

In Section \ref{sec:ergogeo}, we found some null geodesics that cross the horizon.  We now attempt to construct a set of coordinates based around these geodesics, and expect these coordinates to be valid across the horizon.  We hope that this will provide some geometrical insight into why the choice of coordinates across the singly-spinning horizon in \cite{ER:2006} works, and perhaps provide a more convenient set of coordinates for the doubly-spinning case than those of \cite{Elvang:2007bi}.
For convenience, we define functions $\zeta(y)$ and $\xi(x)$, related to the potentials of Section \ref{sec:ergogeo} by \[ \xi(x) \equiv (1-\nu)^2 (-\bxb \Phi^2 - 2\gx \Phi\Psi + \axb \Psi^2 + c \Gx) = -(1-\nu)^2 U(x) \] and  \[ \zeta(y) \equiv (1-\nu)^2( \ayb\Phi^2 - 2\gy \Phi\Psi - \byb \Psi^2 + c \Gy) = -(1-\nu)^2 V(y). \]  Given this, the zero energy, null ergoregion geodesics of Section \ref{sec:geo} are described in our original set of coordinates by \begin{eqnarray*}
\dot{x} &=& \pm \frac{(x-y)^2 (1-\nu)}{R^2 \Hxy}  \sqrt{\xi(x)} \\ 
\dot{y} &=& - \frac{(x-y)^2 (1-\nu)}{R^2 \Hxy} \sqrt{\zeta(y)}  \\
\dot{\phi} &=& \frac{(x-y)^2}{R^2 \Hxy \Gx \Gy}\left[\Axy \Phi - \Lxy \Psi \right] \\
\dot{\psi} &=& \frac{(x-y)^2}{R^2 \Hxy \Gx \Gy}\left[-\Lxy \Phi - \Ayx \Psi \right] \\
\dot{t} &=& -\Omy \dot{\psi} - \Omf \dot{\phi} \\
&=& \frac{(x-y)^2 \la \sqrt{2(1+\nu-\la)(1+\nu+\la)}}{R \Hxy \Hyx \Gx \Gy} \bigg\{ \frac{1+y}{1-\la+\nu}(1+\la-\nu + x^2 y \nu (1-\la-\nu) + \\ 
& & 2\nu x (1-y))[-\Lxy \Phi - \Ayx \Psi ] + (1-x^2)y\sqrt{\nu} [ \Axy \Phi - \Lxy \Psi] \bigg\},
\end{eqnarray*} where we have chosen signs such that $y$ is decreasing with $\tau$; that is we consider the part of a geodesic that is pointing inwards across the horizon.\footnote{We could of course look at the outgoing sections of geodesics by simply changing the sign of the timelike coordinate, which we would expect to produce coordinates suitable for the white hole horizon rather than the black hole.}

Given a geodesic in this class, we might look to find a set of coordinates $(\tau,\tilde{x}^i)$ such that the geodesic is the line $\frac{d}{d\tau}(\tilde{x}^i) = 0$, where $\tau$ is an affine parameter along the geodesic (and $i=1,2,3,4$).  However, a nice feature of the original metric is the symmetry that exists between $x$ and $y$, so attempting to preserve this by transforming only three of the coordinates might well be desirable.  Our revised target will therefore be to find functions $\eta^i(x,y)$ such that, \[ \dot{t} - \frac{\pd \eta^t}{\pd x} \dot{x} - \frac{\pd \eta^t}{\pd y} \dot{y} = \dot{\phi} - \frac{\pd \eta^\phi}{\pd x} \dot{x} - \frac{\pd \eta^\phi}{\pd y} \dot{y} = \dot{\psi} - \frac{\pd \eta^\psi}{\pd x} \dot{x} - \frac{\pd \eta^\psi}{\pd y} \dot{y} =  0.\]  Given this, we can construct the new coordinates $v=t-\eta^t$, $\tilde{\phi} = \phi - \eta^\phi$ and $\tilde{\psi} = \psi - \eta^\psi$.  These three new coordinates will be constant along the geodesics, and therefore we can expect the new coordinate system to be regular at the future horizon.  This is the most general form of coordinate change for these three coordinates that preserves the Killing vectors, that is with \[ \frac{\pd}{\pd v} = \frac{\pd}{\pd t}, \quad \frac{\pd}{\pd \tilde{\phi}} = \frac{\pd}{\pd \phi} \eqand \frac{\pd}{\pd \tilde{\psi}} = \frac{\pd}{\pd \psi}. \]  

\subsection{Singly-spinning case}
To see how this works, we will first apply it to the singly spinning case $\nu=0$.  Here, we have \[ \zeta(y) =c{G(y)}-\Psi^2 {H(y)} \eqand \xi(x) = c{G(x)}-\Phi^2 {H(x)}, \] with equations of motion 
\[ \dot{x} = \pm \frac{(x-y)^2}{R^2 {H(x)}} \sqrt{{\xi(x)}},\quad \dot{\phi} = \frac{(x-y)^2}{R^2 {H(x)}} \left[ \frac{{H(x)}\Phi}{{G(x)}} \right],\]
\[ \quad\dot{y} = - \frac{(x-y)^2}{R^2 {H(x)}} \sqrt{{\zeta(y)}}, \quad \dot{\psi} = \frac{(x-y)^2}{R^2 {H(x)}} \left[ -\frac{{H(y)}\Psi}{{G(y)}} \right], \] and \[ \dot{t} = -\Omy \dot{\psi} = \frac{(x-y)^2}{R^2 {H(x)}} \left[ -\frac{CR(1+y)\Psi}{{G(y)}}\right], \] where the constant $C$ is defined by (\ref{eqn:Cdef}).

Then, \[ \dot{\psi} - \frac{\pd \eta^\psi}{\pd x} \dot{x} - \frac{\pd \eta^\psi}{\pd y} \dot{y} = \frac{(x-y)^2}{R^2 {H(x)}} \left[ -\frac{{H(y)}}{{G(y)}} \mp \sqrt{{\xi(x)}} \frac{\pd \eta^\psi}{\pd x}  + \sqrt{{\zeta(y)}} \frac{\pd \eta^\psi}{\pd y}  \right] .\]   If we pick \[ \eta^\psi = \Psi \int_{y_0}^y \frac{ H(y') dy'}{G(y') \sqrt{\zeta(y')}} \] this vanishes as required.  Similarly, picking \[ \eta^\phi = \pm \Phi \int_{x_0}^x \frac{ H(x') dx'}{G(x') \sqrt{\xi(x')}} \eqand \eta \equiv \eta^t = \Psi \int_{y_0}^y \frac{RC (1+y') dy' }{G(y') \sqrt{\zeta(y')}} \] solves the analogous equations for $\phi$ and $t$.  Note that the lower (constant) bounds $y_0$ and $x_0$ on the integrals above are essentially arbitrary, though care must be taken to make sure that they leave well defined integrals.  A sensible choice, that is guaranteed to be well defined, is to pick $x_0 = 0$, and $y_0$ to be the turning point in the $y$ motion of the geodesic, that is $\zeta(y_0)=0$.  Note that $\pd \eta /\pd y$ and $\pd \eta^\psi /\pd y$ diverge at the horizon.  This is necessary in order to cancel the divergence at the horizon in the original coordinates, and analogous to what happens for coordinate changes across the horizon in more familiar cases.

The resulting change in the basis of 1-forms is\[dv = dt - \frac{CR(1+y)\Psi}{{G(y)} \sqrt{{\zeta(y)}}} dy, \quad d\tilde{\psi} = d\psi - \frac{\Psi {H(y)}}{{G(y)} \sqrt{{\zeta(y)}}} dy \eqand d\tilde{\phi} = d\phi \mp \frac{\Phi {H(x)}}{{G(x)} \sqrt{{\xi(x)}}} dx, \] and this puts the metric (\ref{eqn:singlymet}) into the form \begin{multline} ds^2 = -\frac{{H(y)}}{{H(x)}}(dv +\Omy d\tilde{\psi})^2 + \frac{R^2 {H(x)}}{(x-y)^2} \Bigg[ \frac{c dx^2}{c{G(x)}-\Phi^2 {H(x)}} - \frac{c dy^2}{c{G(y)}-\Psi^2{H(y)}} \\ \pm \frac{2\Phi d\tilde{\phi}dx }{\sqrt{c{G(x)}-\Phi^2 {H(x)}}} -  \frac{2\Psi  d\tilde{\psi}dy }{\sqrt{c{G(y)}-\Psi^2 {H(y)}}}+  \frac{{G(x)}}{{H(x)}} d\tilde{\phi}^2 - \frac{{G(y)}}{{H(y)}} d\tilde{\psi}^2 \Bigg] .\end{multline}  This nicely preserves the $x \leftrightarrow y$, $\phi \leftrightarrow \psi$ symmetry of the original metric.  The inverse metric is given by \begin{multline} \label{eqn:newsinginv} \left( \frac{\pd}{\pd s} \right)^2 = -\frac{{H(x)}}{{H(y)}} \left( \frac{\pd}{\pd v} \right)^2 + \frac{(x-y)^2}{R^2 {H(x)}} \Bigg[ {G(x)} \left( \frac{\pd}{\pd x} \right)^2 -  {G(y)} \left( \frac{\pd}{\pd y} \right)^2 \\ \mp \frac{ 2\Phi {H(x)}}{ \sqrt{{\xi(x)}}}  \frac{\pd}{\pd \tilde{\phi}}  \frac{\pd}{\pd x}  + \frac{ 2\Psi {H(y)}}{ \sqrt{{\zeta(y)}}} \left( \frac{\pd}{\pd \tilde{\psi}} - \Omy \frac{\pd}{\pd v}\right)  \frac{\pd}{\pd y} + c \frac{{H(x)}}{{\xi(x)}} \left( \frac{\pd}{\pd \tilde{\phi}}  \right)^2  - c \frac{{H(y)}}{{\zeta(y)}} \left( \frac{\pd}{\pd \tilde{\psi}} - \Omy \frac{\pd}{\pd v} \right)^2 \Bigg] . \end{multline}  Since the components of both the metric, and its inverse are regular at $y=y_h$, it is now a well defined coordinate system across the horizon $G(y)=0$.  Note that, like in the original form of the metric, there is a coordinate singularity as we approach $x=\pm 1$, which has no physical significance, and simply corresponds to the singularity at the origin of plane polar coordinates.  There is a further subtlety here though, since we saw in \S\ref{sec:ergogeo} that for $|\Phi|>0$ the allowed range of $x$ along the geodesic is limited (since ${\xi(x)}<0$ for some $x\in [-1,1]$).  Thus, these coordinates can only cover the entire horizon when we set $\Phi=0$.

The simplest geodesics discussed in \S \ref{sec:ergogeo}  were those with $c=0=\Phi^2$, and $\Psi^2 > 0$.  This leaves us with the transformation \[ dv = dt - \frac{CR(1+y)}{{G(y)}\sqrt{-{H(y)}}} dy, \quad d\tilde{\psi} = d\psi + \frac{\sqrt{-{H(y)}}}{{G(y)}} dy \eqand d\tilde{\phi} = d\phi \] which is precisely the coordinate change given in \cite{ER:2006}, leaving the metric in the form \[ ds^2 = -\frac{{H(y)}}{{H(x)}}(dv +\Omy d\tilde{\psi})^2 + \\ \frac{R^2 {H(x)}}{(x-y)^2} \left[\frac{dx^2}{G(x)} +  \frac{{G(x)}}{{H(x)}} d\tilde{\phi}^2 - \frac{2 d\tilde{\psi}dy }{\sqrt{- {H(y)}}} - \frac{{G(y)}}{{H(y)}} d\tilde{\psi}^2 \right] .\] Thus, this technique has generated a family of possible coordinate transformations, including those that are already known, and attached a geometric significance to them.  Note that the coordinates are only valid out as far as the turning point $y=y_0$ of the geodesic in question, that is for the region $-\infty < y < y_0$ where $\zeta(y_0)=0$.  There is still a coordinate singularity at $x=\pm 1$, as in the original set of coordinates.

Note that, if we wished, it would be possible to make a further change of coordinates $x\mapsto \tilde{x}$ such that $\dot{\tilde{x}}=0$ along the geodesics.  However, the range of the new coordinate $\tilde{x}(x,y)$ is messy (and $y$ dependent), and the resulting form of the metric is unpleasant and of no obvious practical use.  Having done this, $y$ is the only coordinate varying along the geodesics, and it does so monotonically if we only consider the ingoing part of the geodesic (as we have been doing).  Thus, we can write $x=x(y)$ along the geodesic, and hence use the affine parameter $\tau$ rather than $y$ as the remaining coordinate, leaving us with the type of coordinate system originally suggested above.  We do not present any of this explicitly here, since the resulting form of the metric is messy in the extreme, and not obviously of any real value.

\subsection{Doubly-spinning case}
Now we move on to the doubly spinning case.  Here the form of the geodesic equations is more complicated, so we expect the coordinate change associated with it to be more complicated as a result.  We need to solve the PDEs \[ \dot{\phi} - \frac{\pd \eta^\phi}{\pd x} \dot{x} - \frac{\pd \eta^\phi}{\pd y} \dot{y} = 0 \eqand \dot{\psi} - \frac{\pd \eta^\psi}{\pd x} \dot{x} - \frac{\pd \eta^\psi}{\pd y} \dot{y} = 0 \] which can be written as \[ \frac{(x-y)^2}{R^2 \Hxy} \left[ \left(\frac{\bxb \Phi+\gx \Psi}{\Gx} \mp \sqrt{\tzx} \frac{\pd \eta^\phi}{\pd x} \right)  + \left(\frac{\ayb \Phi - \gy \Psi}{\Gy} + \sqrt{\zy} \frac{\pd \eta^\phi}{\pd y} \right)  \right] = 0 \] and  \[  \frac{(x-y)^2}{R^2 \Hxy} \left[ \left(\frac{\gx\Phi - \axb \Psi}{\Gx} \mp \sqrt{\tzx} \frac{\pd \eta^\psi}{\pd x} \right)  + \left(\frac{-\gy\Phi - \byb \Psi}{\Gy} + \sqrt{\zy} \frac{\pd \eta^\psi}{\pd y} \right)  \right] = 0 . \]  They have the obvious separable solutions \begin{equation} \label{eqn:etaf} \eta^\phi = \pm \int_{x_0}^x \frac{{\beta}(x') \Phi + \gamma(x') \Psi}{G(x') \sqrt{\xi(x')}} dx' + \int_{y_0}^y \frac{ -{\alpha}(y')\Phi + \gamma(y') \Psi}{G(y') \sqrt{\zeta(y')}} dy'\end{equation} and \begin{equation} \label{eqn:etay} \eta^\psi = \pm \int_{x_0}^x \frac{ \gamma(x') \Phi - {\alpha}(x') \Psi}{G(x') \sqrt{\xi(x')}} dx' + \int_{y_0}^y \frac{ \gamma(y') \Phi + {\beta}(y') \Psi}{G(y') \sqrt{\zeta(y')}} dy'\end{equation}

However, it is less easy to solve the PDE \begin{equation}\label{eqn:tdot} \dot{t} - \frac{\pd \eta}{\pd x} \dot{x} - \frac{\pd \eta}{\pd y} \dot{y} =0,\end{equation} since the dependence of $\Omf$ and $\Omy$ on both $x$ and $y$ means that the equation does not separate.  In order to get a new set of coordinates that is analogous to that of the singly spinning case, we might hope to be able to set $v = t - \eta^t$ where \begin{equation} d\eta^t = -\Omy d\eta^\psi - \Omf d\eta^\phi,\label{eqn:nogood} \end{equation} which would give us the convenient result $dt+\Omf d\phi +\Omy d\psi = dv +\Omf d\tilde{\phi} +\Omy d\tilde{\psi}$.  Unfortunately, the right hand side of (\ref{eqn:nogood}) is not a total derivative for $\nu>0$, so this is impossible.

Instead, we might take either of two different approaches:\begin{itemize} \item Look for an exact solution of (\ref{eqn:tdot}), even if it cannot be written in a convenient separable form like (\ref{eqn:etaf}), (\ref{eqn:etay}). 
\item Give up on completely solving (\ref{eqn:tdot}), and instead just look for an $\eta$ such that $v=t-\eta$ has \begin{equation} \dot{v} = \dot{t} - \frac{\pd \eta}{\pd x} \dot{x} - \frac{\pd \eta}{\pd y} \dot{y} < \infty \label{eqn:tfinite} \end{equation} at the horizon $y=y_h$, as we move along one of the geodesics.\end{itemize}  We have investigated both of these possibilities.

\subsubsection{Construction of an exact solution to (\ref{eqn:tdot})}
It is possible to solve equation (\ref{eqn:tdot}) exactly, using the method of characteristics.  However, this solution turns out to be fairly complicated, and as such is not particularly useful for constructing a new set of coordinates.  The construction of this solution is described in Appendix \ref{app:tdot}.  The end result is that we get functions in the new metric that, though well defined, can only be written down implicitly in terms of inverse functions, which would make the resulting metric highly inconvenient to work with.  Furthermore, regularity at the horizon is not manifest, which is the main motivation for doing this.

\subsubsection{Solving the finiteness condition only}
Instead, we look for a new time coordinate $v$ that is merely finite at the horizon, rather than constant everywhere along the geodesic.  As described above, Elvang and Rodriguez \cite{Elvang:2007bi} showed how to construct coordinates (\ref{eqn:elvrodcoords}) that are valid across the horizon.  It is useful to pause for a moment to understand how their change of coordinates works, since it will be useful in constructing a suitable $v$ here.

In order for the coordinate system across the horizon to be well-defined, we require that the divergence in $g_{yy}$ has been removed by this coordinate change, and that no new divergences are introduced in any of $g_{\bar{t} y}$, $g_{\bar{\phi} y}$ or $g_{\bar{\psi} y}$.  A straightforward computation shows that these conditions are equivalent to requiring that $A$,$B$,$C$ can be chosen such that, for all $x$, \begin{eqnarray}
C+\Omf(x,y_h) A + \Omy(x,y_h) B &=& 0 \label{eqn:hor1} \\
A(y_h,x) A - L(x,y_h) B &=& 0 \label{eqn:hor2} \\
-L(x,y_h) A - A(x,y_h) B &=& 0 \label{eqn:hor3}\\
\lim_{y\rightarrow y_h} \Bigg[ - \frac{\Hxy}{\Gy} + \frac{A}{y-y_h} \left( \frac{\Ayx A - \Lxy B }{ \Hyx (y-y_h)} \right) + \;\;\;\;\;\;\;\;\;\;\;\;\;\;\;\;\;\;\;\;\;\;\;\;\;\;\;\;\;\;   && \nonumber \\ 
\frac{B}{y-y_h} \left( \frac{-\Lxy A - \Axy B }{ \Hyx (y-y_h)} \right) \Bigg] &<& \infty \label{eqn:hor4}. \end{eqnarray}

It is not immediately obvious that it is possible to satisfy these conditions simultaneously, though of course it must be if the doubly-spinning black ring is a well defined black hole spacetime.  Expanding in $x$ shows that (\ref{eqn:hor1}),(\ref{eqn:hor2}),(\ref{eqn:hor3}) have a 1-parameter family of solutions given by \begin{equation}
\frac{C}{B} = R\sqrt{2\left(\frac{1+\nu+\la}{1+\nu-\la} \right)} \eqand \frac{A}{B} = -\frac{\sqrt{\nu} (1+y_h^2)}{\la y_h} =\frac{\sqrt{\nu}}{\la} [\la - y_h (1-\nu)] =\frac{ \gamma(y_h)}{{\beta}(y_h)} = -\frac{{\alpha}(y_h)}{\gamma(y_h)}.
\end{equation} Putting this into (\ref{eqn:hor4}) fixes $B$, and hence $A$ and $C$.  Note that carrying out this last step explicitly is very fiddly, and its validity relies on the non-trivial fact that \[ H(x,y_h) H(y_h,x) = \mathrm{const} . \left[ (A/B)^2 \axb + 2(A/B) \gx - \bxb \right] \] for some constant, where $A/B$ is as given above.

How does this link in to our solutions above?  We will see below that our change in coordinates makes the metric finite at the horizon, and hence it can only differ from the coordinate change of \cite{Elvang:2007bi} by a finite amount, that is as $y\rightarrow y_h$, \begin{equation} (y-y_h) \frac{\pd \eta^\phi}{\pd y} \rightarrow A \eqand (y-y_h) \frac{\pd \eta^\psi}{\pd y} \rightarrow B .\label{eqn:horlims} \end{equation}  Explicit computation confirms that this is the case.  Furthermore, we will see below that our change of coordinates $(\phi,\psi) \rightarrow (\tilde{\phi},\tilde{\psi})$ renders the $R^2/(x-y)^2$ part of the line element finite for any choice of $v$, and hence we do not need to do the fiddly computation to work the value of $B$ using (\ref{eqn:hor4}), but can merely read it off from (\ref{eqn:horlims}), that is \[ B = \lim_{y\rightarrow y_h}\left[(y-y_h) \frac{\pd \eta^\psi}{\pd y} \right] = \frac{\gamma(y_h) \Phi + {\beta}(y_h)\Psi}{(1-y_h^2)\sqrt{(\la^2-4\nu)\zeta(y_h)}}. \]  This is a significantly easier approach for getting this result.

Given this, we can immediately see that a valid change of time coordinate, to render the metric finite in the non-extremal case, is to set \[ dv = dt - \frac{C}{y-y_h}dy,\quad \mathrm{where} \quad C = R\sqrt{2\left(\frac{1+\nu+\la}{1+\nu-\la} \right) } \frac{\gamma(y_h) \Phi + {\beta}(y_h)\Psi}{(1-y_h^2)\sqrt{(\la^2-4\nu)\zeta(y_h)}}. \]  This can be made slightly neater if we write \[ dv = dt - \frac{D (\gamma(y) \Phi + {\beta}(y)\Psi) }{G(y)\sqrt{\zeta(y)}}dy \quad \mathrm{where} \quad D = R \sqrt{2\left(\frac{1+\nu+\la}{1+\nu-\la}\right) },\] which has the correct limit at the horizon, and will allow the new metric to be written more conveniently.

This transforms the first part of the metric via \[ dt + \Omf d\phi + \Omy d\psi = dv + \Omf d\tilde{\phi} + \Omy d\tilde{\psi} + \tilde{\Om}_x dx + \tilde{\Om}_y dy \equiv dv + \tilde{\Omega} \] where \begin{equation} \tilde{\Om}_x = \pm \frac{ \Phi (\Omf \bxb + \Omy \gx) + \Psi (\Omf \gx - \Omy \axb)}{\Gx \sqrt{\tzx}} \end{equation} and \begin{equation} \tilde{\Om}_y = \frac{ \Phi (D \gy - \Omf \ayb - \Omy \gy) + \Psi (D \byb + \Omf \gy + \Omy \byb)}{\Gy \sqrt{\zy}}. \label{eqn:Omyt} \end{equation} These are fairly complicated, but are suitably regular as we approach the horizon (though this regularity is not immediately manifest from looking at (\ref{eqn:Omyt}).  Furthermore, they remain valid in the extremal limit, while the original approach of \cite{Elvang:2007bi} needs additional corrections in this case.

\subsubsection{Transformed metric} \label{sec:transmet}
Given the above form for $ \tilde{\Om} $, we find that the metric can be written in the new coordinates as \begin{multline} \label{eqn:newmet} ds^2 = -\frac{\Hyx}{\Hxy} ( dv+ \tilde{\Om} )^2 + \\
 \frac{R^2 \Hxy}{(x-y)^2 (1-\nu)^2} \Bigg[ \left(\frac{\tzx}{\Gx} + \frac{\Ayx\tx^2 - 2\Lxy \tx \cx - \Axy \cx^2}{\Gx ^2 \Hxy\Hyx} \right) d(x,y)^2 + \\
2\left( -c (1-\nu)^2 \frac{dy}{\sqrt{\zy}} + \frac{(\Ayx \tx - \Lxy \cx) d\tilde{\phi}- (\Lxy \tx + \Axy \cx) d\tilde{\psi}}{\Gx \Hxy \Hyx} \right)  d(x,y) \\
- 2(1-\nu)^2 (\Phi d\tilde{\phi} + \Psi d\tilde{\psi}) \frac{dy}{\sqrt{\zy}} + \frac{\Ayx d\tilde{\phi}^2-2\Lxy d\tilde{\phi} d\tilde{\psi}- \Axy d\tilde{\psi}^2}{\Hxy \Hyx} \Bigg] , \end{multline} where \begin{equation} d(x,y) \equiv  \left( \pm\frac{dx}{\sqrt{ \tzx}} + \frac{dy}{\sqrt{\zy}} \right), \quad \theta(x)  \equiv  {\beta}(x) \Phi + \gamma(x) \Psi \eqand \chi(x) \equiv \gamma (x) \Phi - {\alpha}(x) \Psi . \end{equation}

In the singly spinning case we were able to maintain the $x\leftrightarrow y$ symmetry after the change of coordinates, but this turns out to not be possible here if we want to write the metric in a manner that is manifestly well defined as we cross the horizon.  As a result, this form of the metric is somewhat unpleasant.  Note that it has the following properties: \begin{itemize}
\item The metric (and also its inverse) are regular at the horizon $y=y_h$.
\item There is still a coordinate singularity at $x=\pm 1$.
\item It depends on three arbitrary parameters $c$, $\Phi$ and $\Psi$, any two of which are independent.
\end{itemize}

As in the singly spinning case, we have found a family of geodesics with two free parameters (any two of $c$, $\Psi$, $\Phi$), so we are free to pick their values so as to simplify the metric in order to find something that might be more useful for practical applications.  As in the singly spinning case, $\Phi=0$ is a natural, legitimate choice, but unfortunately we can no longer set $c=0$ (see Section \ref{sec:dsgeo}).   Also, as in the singly spinning case, the coordinates have a restricted $x$ range for $\Phi \neq 0$.

The line element in the $\Phi=0$ case can be written in the form \begin{multline} \label{eqn:newmet2} ds^2 = -\frac{\Hyx}{\Hxy} ( dv + \tilde{\Om} )^2 + \frac{R^2 \Hxy}{(x-y)^2 (1-\nu)^2} \Bigg[\bar{c} \left( \frac{dx^2}{\bar{c} \Gx + \axb} - \frac{dy^2}{\bar{c} \Gy - \byb} \right)+ \\
 \left(\frac{\axb}{\Gx} + \frac{\Ayx\gx^2 + 2\Lxy \gx \axb - \Axy \axb^2}{\Gx ^2 \Hxy\Hyx (1-\nu)^2} \right) d(x,y)^2+ \\
2\left(\frac{(\Ayx \gx + \Lxy \axb) d\tilde{\phi} - (\Lxy \gx - \Axy \axb) d\tilde{\psi}}{\Gx \Hxy \Hyx} \right) d(x,y) \\ 
\frac{- 2(1-\nu) d\tilde{\psi} dy}{\sqrt{\bar{c} \Gy - \byb}} + \frac{\Ayx d\tilde{\phi}^2-2\Lxy d\tilde{\phi} d\tilde{\psi}- \Axy d\tilde{\psi^2}}{\Hxy \Hyx} \Bigg], \end{multline} where now \[ d(x,y) \equiv \left( \pm\frac{dx}{\sqrt{\bar{c} \Gx + \axb}} + \frac{dy}{\sqrt{\bar{c} \Gy - \byb}} \right). \]  This now contains only the one arbitrary constant $\bar{c} \equiv c/\Psi^2$.  At first glance this looks equally complicated, but the only polynomial functions that appear in this expression are now those that appear in the original metric itself, so progress has been made.

From Section \ref{sec:dsgeo} we have the condition that \[ \bar{c} \geq \frac{\nu}{1+\nu-\la} \left[ 2(1+\la) - 3\la\nu + \nu(1-\nu) \right] \] for these coordinates to be valid for all $x$ (with the exception of the coordinate singularity on the axis $x=\pm 1$.  We might hope that by saturating this bound we could obtain a simpler form for the metric (as occurs in the $c=0$ case for the singly spinning ring), but it is far from clear that this is the case.  Of course doing so does remove the last arbitrary constant from the metric and thus fix it entirely, as well as providing what seems like the natural doubly spinning generalisation of the singly spinning result of \cite{ER:2006}.  It would be interesting to see if a value of $\bar{c}$ could be chosen that really simplified things further here, but we have been unable to do this successfully.

It seems that no further progress can be made in our study of coordinate systems, so finally we move on to discuss whether the separability of part of the HJ equation that we have discovered can be used to say anything about hidden symmetries of the spacetime.

\section{Hidden Symmetries} \label{sec:sym}
If a $d$-dimensional metric has at least $d-1$ commuting Killing vectors, corresponding to $d-1$ Noether symmetries then its associated Hamilton-Jacobi equation has separable solutions.  On the other hand, if it has fewer Killing vectors, but its HJ equation is still separable, then this separability can be linked to a hidden phase space symmetry, related to the existence of a higher-rank Killing tensor $K$ satisfying the generalised Killing equation \[ \nabla_{(a} K_{b_1 b_2...b_p)} = 0 .\]  In most known cases, this tensor is rank-2.  These tensors have been studied for many other black hole spacetimes, beginning with the Kerr black hole \cite{Walker:1970}.

Separability in the null case only is a conformally invariant property of the geometry, while the Killing equation is not conformally invariant.  There is a conformally invariant generalisation though, which in the rank-2 case reads \begin{equation} \nabla_{(a} K_{bc)}= \omega_{(a} g_{bc)} \label{eqn:confkilling} \end{equation} for some 1-form $\omega$, given in dimension $d$ by \[ \omega_a = \frac{2}{d+2} \left[ \nabla^b K_{ab} + \frac{1}{2} \nabla_a (\mathrm{tr} K) \right].\]  If $K_{ab}$ solves this for a spacetime $(M,g)$, then $\La^4 K_{ab}$ solves it for the conformally related spacetime $(M,\La^2 g)$ for any suitably regular function $\La^2$.  Solutions of this equation are referred to as \emph{conformal Killing} (CK) tensors, and they have the property that $K^{ab} p_a p_b$ is conserved along any null geodesic with momentum $p_a$.

Note that the metric is itself a Killing tensor, with associated conserved quantity $\mu^2$, the mass of a particle following a geodesic.  We say that a rank-2 CK tensor is \emph{irreducible} (or \emph{non-trivial}) if it cannot be expressed in terms of the metric $g$ and Killing vectors $\{ k^{(i)} \}$ in the form \[ K_{ab} = a(x^c) g_{ab} + \sum_{i,j} b^{ij} k_{(a}^{(i)} k_{b)}^{(j)}, \] for some scalar function $a(x^c)$ and constants $b^{ij}$.  Two CK tensors are \emph{independent} if their difference is irreducible.  Other recent work on CK tensors includes \cite{Chow:2008} and \cite{Hioki:2008}.

A metric with $d-2$ mutually commuting Killing vectors can be written in a form where its components depend on only two coordinates, $x$ and $y$ say.  Then, if the HJ equation is separable for null geodesics, it can be written in the form \[ K^{ab}_{(1)}(x) p_a p_b = K^{ab}_{(2)}(y) p_a p_b = \mathcal{K}  \] for some constant $\mathcal{K}$.  Both $K_{(1)}$ and $K_{(2)}$ must be CK tensors for the geometry, and they satisfy the relation \begin{equation}K^{ab}_{(1)}(x) - K^{ab}_{(2)}(y) = f(x,y) g^{ab} \label{eqn:KminusK}\end{equation} for some function $f(x,y)$.  Therefore, they are not independent.

Does anything similar apply for the black ring metric?  We have a separable form (\ref{eqn:sep}) for the HJ equation, but only in the null, zero energy case.  We can read off tensors $K_{(1)}$ and $K_{(2)}$ from this, but do not expect them to be conformal Killing tensors, due to the $E=0$ condition.  Note that the components $K^{tt}$ and $K^{ti}$ of these tensors appear somewhat arbitrary, since they do not have any effect on the value of \[ \frac{c}{(1-\nu)^2} = K^{ab} p_a p_b = K^{tt} E^2 -2 K^{ti} E p_i + K^{ij} p_i p_j = K^{ij} p_i p_j\] along one of the separable geodesics.  This hints at a way of understanding the symmetry that allows for this separation; dimensional reduction to remove the $K^{ta}$ components. This turns out to be a neat way of dealing with the zero-energy condition on these geodesics.

\subsection{Kaluza-Klein Reduction}
We perform a dimensional reduction to project out the $\pd/\pd t$ direction, via the standard Kaluza-Klein procedure.  We take an ansatz \[ ds^2 = e^{\varphi/\sqrt{3}} h_{ij} dx^i dx^j + e^{-2\varphi/\sqrt{3}} (dt + \mathcal{A}_i dx^i) \] where $i,j,...$ range over $x,\phi,y,\psi$ and $h_{ij}$ is the metric on the 4-dimensional space.  
Note that $\pd/\pd t$ is spacelike in the ergoregion (to which our known geodesics are restricted), so the reduced metric has signature $(+,+,+,-)$, and we must restrict the ranges of our coordinates in the reduced metric so that they only correspond to this region (otherwise we would be performing a timelike reduction, which would require a slightly different analysis).  It is well known that the resulting 4-dimensional geometry solves the Einstein-Maxwell-Dilaton equations.

Comparison to the line element (\ref{eqn:metric}) gives \[ e^{-2\varphi / \sqrt{3} } = -\frac{H(y,x)}{H(x,y)} \; \; \mathrm{and} \; \; \mathcal{A}_i dx^i = \Om = \Om_\phi d\phi + \Om_\psi d\psi .\]  Given this, it is straightforward to show that the dimensionally reduced metric is given by \begin{equation} \label{eqn:4met} ds_4^2 = h_{ij} dx^i dx^j =  \La^2(x,y) \left[ \frac{dx^2}{G(x)}-\frac{dy^2}{G(y)}+\frac{A(y,x) d\phi^2-2L(x,y) d\phi d\psi- A(x,y) d\psi^2}{H(x,y) H(y,x)} \right] \end{equation} where \begin{equation}\label{eqn:Tnat} \La^2(x,y) \equiv \frac{R^2 \sqrt{-H(x,y)H(y,x)}}{(x-y)^2 (1-\nu)^2}.\end{equation}

Note that the singly-spinning black ring was originally constructed in \cite{ER:2001} by analytic continuation of an oxidised Kaluza-Klein C-metric \cite{Chamblin:1996}.  Here, we have found a Kaluza-Klein metric of a similar form to the C-metric that is linked more directly to the black ring; that is to say no analytic continuation is required.  Furthermore, this reduction is equally valid in the doubly-spinning case, for which a C-metric associated with the ring does not exist in the literature.

\subsection{Conformal Killing Tensors} \label{sec:confkill}
Note that the zero-energy geodesics in the 5-dimensional metric correspond precisely to the geodesics of the 4-dimensional metric (while those which are not zero-energy are related to charged particle orbits).  In the 5 dimensional case we know all of the zero energy, null geodesics, so this translates to knowing all of the null geodesics in the 4 dimensional metric.  Therefore, as described above, we should expect that the dimensionally reduced metric has a CK tensor, and now proceed to show that this is indeed the case.

In order to see the conformal invariance explicitly, it is nice to do the calculation with a general conformal factor $\La^2=\La^2(x,y)$ in the metric (\ref{eqn:4met}), where of course equation (\ref{eqn:Tnat}) gives the choice of $\La^2$ that actually results from the Kaluza-Klein reduction of the black ring.

We read off the forms of $K_{(1)}^{ij}$ and $K_{(2)}^{ij}$ from (\ref{eqn:sep}), which gives non-vanishing components \begin{equation} \begin{array}{cc}
K_{(1)}^{xx} = G(x)                                      , & K_{(2)}^{yy} = G(y), \\& \\
K_{(1)}^{\phi\phi} = \frac{{\beta} (x)}{(1-\nu)^2 G(x)}  , & K_{(2)}^{\phi\phi} = \frac{- {\alpha} (y)}{(1-\nu)^2 G(y)}  \\& \\
K_{(1)}^{\phi\psi} = \frac{\gamma (x)}{(1-\nu)^2G(x)}    , & K_{(2)}^{\phi\psi} = \frac{\gamma (y)}{(1-\nu)^2 G(y)},\\& \\
K_{(1)}^{\psi\psi} = \frac{-\alpha (x)}{(1-\nu)^2 G(x)}  , & K_{(2)}^{\psi\psi} = \frac{{\beta} (y)}{(1-\nu)^2 G(y)}.\end{array} \end{equation} Now \[ K_{(1)}^{ij} - K_{(2)}^{ij} = \La^2 h^{ij} ,\] so if one of these tensors is a conformal Killing tensor, so is the other, and they are not independent.  Given this, perhaps the natural choice of CK tensor to work with is $K \equiv  K_{(1)} + K_{(2)}$.

Differentiating, we see that $K$ satisfies the conformal Killing equation \[ \nabla_{(i} K_{jk)} = \omega_{(i} h_{jk)} \quad \mathrm{where} \quad \omega = 2\La \left[ \frac{\pd \La}{\pd x} dx -  \frac{\pd \La}{\pd y} dy \right], \]  and is therefore a CK tensor.  Note that $K$ is actually a Killing tensor of the geometry that has constant conformal factor $\La^2$.

With indices raised, $K^{ij}$ is not dependent on the conformal factor, and with coordinates $(x,\phi,y,\psi)$, it can be written in matrix form as \begin{equation} \label{eqn:matK} \mathbf{K} =  \left( \begin{array}{cccc} G(x) & 0 & 0 & 0 \\
0 & \frac{1}{(1-\nu)^2} \left( \frac{\bxb}{\Gx} - \frac{\ayb}{\Gy} \right) & 0 & \frac{1}{(1-\nu)^2} \left( \frac{\gx}{\Gx} + \frac{\gy}{\Gy} \right)\\ 0 & 0 & G(y) & 0 \\
0   & \frac{1}{(1-\nu)^2} \left( \frac{\gx}{\Gx} + \frac{\gy}{\Gy} \right) & 0 & \frac{1}{(1-\nu)^2} \left( \frac{\byb}{\Gy} - \frac{\axb}{\Gx} \right) \end{array}\right) . \end{equation}

There is an alternative way of seeing the existence of this conformal Killing tensor.  Benenti and Francaviglia \cite{Benenti:1979} give a canonical form for the metric of an $n$-dimensional spacetime admitting $(n-2)$ Killing vectors, and a non-trivial rank-2 Killing tensor.  The inverse metric takes the form \begin{multline} g^{-1} = \frac{1}{\varphi_1(x^1) + \varphi_2(x^2)} \Bigg[ \psi_1(x^1) \left( \frac{\pd}{\pd x^1} \right)^2 + \psi_2(x^2) \left( \frac{\pd}{\pd x^2} \right)^2 + \\
\left(\psi_1(x^1) \zeta_1^{\alpha\beta}(x^1) + \psi_2(x^2) \zeta_2^{\alpha\beta}(x^2) \right) \left( \frac{\pd}{\pd \phi^\alpha} \right)\left( \frac{\pd}{\pd \phi^\beta} \right) \Bigg] \end{multline} for some functions $\psi_a(x^a)$, $\varphi_a(x^a)$, $\zeta_a^{\alpha\beta}(x^a)$ depending on a single coordinate only, with $ \varphi_1 \psi_1^2 +  \varphi_2 \psi_2^2 \neq 0$ everywhere.  The indices $\alpha, \beta=3,...,n$ label the Killing directions $\pd/\pd \phi^\alpha$.  The rank-2 Killing tensor is given by \[ K^{\alpha \beta} = \frac{1}{\varphi_1 + \varphi_2}\left( \zeta_1^{\alpha\beta} \psi_1 \varphi_2 - \zeta_2^{\alpha\beta} \psi_2 \varphi_1 \right)  , \quad K^{11} = \frac{\varphi_2 \psi_1 }{\varphi_1 + \varphi_2} \eqand K^{22} = \frac{-\varphi_1 \psi_2 }{\varphi_1 + \varphi_2}. \]

The inverse metric for the dimensionally reduced black ring is conformally related to a metric of this form, with $\varphi_a \equiv 1$ and we must therefore have a rank-2 conformal Killing tensor.  The form for this given corresponds precisely to our tensor $K^{ij}$, up to an arbitrary constant factor.

\subsection{Conformal Killing-Yano Tensors}\label{sec:cky}
Often, a conformal Killing (CK) tensor can be constructed from a more fundamental object, a conformal Killing-Yano (CKY) tensor, that is a 2-form $k$ satisfying the conformal Killing-Yano equation \[ \nabla_{(a} k_{b) c} = g_{ab} \xi_c - \xi_{(a} g_{b)c} \quad \mathrm{where} \quad \xi_b = \frac{1}{d-1} \nabla^a k_{ab}.\]  Note that if $k_{ab}$ solves it for spacetime $(M,g)$, then $\La^3 k_{ab}$ solves it for $(M,\La^2 g)$.  Given a CKY tensor $k$, $K_{ab} = k_{ac}k_b^{\phantom{b}c}$ is a CK tensor.  In this case, it turns out that a CKY tensor exists if and only if the ring is singly spinning.

\subsubsection{Singly Spinning Case}
In the singly spinning case, it is straightforward to directly construct an antisymmetric tensor that squares to the Killing tensor $K^{ij}$, that is a $k^{ij}$ such that $ K^{ij} = k^{ik} k^{jl} h_{kl} . $  The tensor \[ k^{x\phi} = \frac{\sqrt{H(x)}}{\La(x,y)} = -k^{\phi x} \eqand k^{y\psi} = \frac{\sqrt{-H(y)}}{\La(x,y)} = -k^{\psi y} , \]  with all other components vanishing, satisfies this.  Lowering indices, this gives us a 2-form \[ k = \La^3 \left[\frac{1}{\sqrt{H(x)}} dx \wedge d\phi - \frac{1}{\sqrt{-H(y)}} dy \wedge d\psi \right] .\]  Note that there is a second tensor with the same property, which can be obtained by taking the Hodge dual of $k$, resulting in \[ \star k = \La^3 \left[\frac{1}{\sqrt{H(x)}} dx \wedge d\phi + \frac{1}{\sqrt{-H(y)}} dy \wedge d\psi \right] .\] By explicit calculation, it can be shown that \[ \nabla_{i} k_{jk} = \nabla_{[i} k_{jk]} + 2 h_{i [j} \xi_{k]} \quad \mathrm{where} \quad \xi = \frac{\Gx}{\sqrt{\Hx}} \frac{\pd \La}{\pd x} d\phi + \frac{\Gy}{\sqrt{-\Hy}} \frac{\pd \La}{\pd y} d\psi \] and therefore $k$ satisfies the conformal Killing-Yano equation (as does $\star k$).

It is interested to briefly consider the case of constant $\La^2$, although this does not correspond to the actual dimensional reduction of the black ring.  Here, $k$ is a Killing-Yano tensor, and its square is a Killing tensor.  In fact something stronger can be said.  Recently, Krtou\v s, Frolov and Kubiz\v n\' ak \cite{Krtous:2008} (strengthening a result of Houri, Oota and Yasui \cite{Houri:2008}) have shown that any $d$-dimensional spacetime manifold with a globally defined closed CKY tensor $k$ (known as a \emph{principal} CKY tensor) can be written in a particular canonical form.  If the geometry solves the vacuum Einstein equations (possibly with a cosmological constant), then this form reduces to the Kerr-NUT-(A)dS metric \cite{Chen:2006}.

Here, taking an exterior derivative gives that \[ dk = -3 \La^2  dx \wedge dy \wedge \left[ \frac{\pd \La}{\pd y} \frac{d\phi}{\sqrt{\Hx}} + \frac{\pd \La}{\pd x} \frac{d\psi}{\sqrt{-\Hy}} \right] \] and hence we see that $k$ is closed for the 4-geometry with constant $\La^2$ (as is $\star k$).  Thus we have a principal CKY tensor here.  The existence of this tensor implies that the metric can be written in the known canonical form, separability of the HJ equation for all geodesics (rather than just null ones), as well as that this 4-metric is of algebraic type $D$.  Since the algebraic type of a metric is a conformally invariant property, the 4-dimensional geometry must be type $D$ for all choices of conformal factor, and therefore the geometry that results directly from the KK reduction of the singly-spinning ring is also type $D$.

\subsubsection{Doubly Spinning Case}\label{sec:doublycky}
In the doubly spinning case, it turns out that the conformal Killing tensor $K^{ij}$ is not derivable from a conformal Killing-Yano tensor.  Furthermore, this result is independent of our particular choice of CK tensor, and therefore proves that no CKY tensor can exist for the doubly-spinning ($\nu>0$) metric.

\begin{lemma}
Define a symmetric rank-(2,0) tensor $K'$ by \[K' = K + C(x^k) h^{-1} + 
p \left( \frac{\pd}{\pd \phi} \right)^2 + 
2 q \left( \frac{\pd}{\pd \phi}\right) \left(\frac{\pd}{\pd \psi} \right) +
r \left( \frac{\pd}{\pd \psi}\right)^2 .\]  Then $K'$ has the following properties:\begin{enumerate}
\item It is a conformal Killing tensor for all differentiable functions $C(x^k)$, and constants $p$, $q$, $r$.
\item Up to arbitrary constant rescalings of $K$, it is the most general irreducible CK tensor.
\item For $\nu>0$, and for any $C(x^a)$, $p$, $q$, $r$, there does not exist an antisymmetric tensor $k$ such that \begin{equation} \label{eqn:fsquare} K'^{ij} = k^{ik} k^{jl} h_{kl}.\end{equation} \end{enumerate} \end{lemma}  Note that if $k$ is a CKY tensor, then a $K'$ defined by (\ref{eqn:fsquare}) must be a CK tensor, and therefore the non-existence of a square-root for the most general non-trivial CK tensor proves the non-existence of a CKY tensor.  Thus, as a direct corollary of this lemma, we see that the dimensional reduction of the black ring spacetime possesses a CKY tensor if and only if the ring is singly-spinning.  When one CKY tensor exists, a second can be constructed by taking the Hodge dual, as described above.  A direct proof of the lemma is given in Appendix \ref{app:symproof}.

\subsection{Klein-Gordon Equation}\label{sec:kg}
Often, when a spacetime possesses a Killing tensor, it is possible to find multiplicatively separable solutions of the Klein-Gordon (KG) equation.  Here, we have additive separability for geodesic motion in the null, zero energy case, so we might hope that this would translate into being able to find time-independent separable solutions to the massless KG equation for the 5-dimensional black ring.  However, the results linking the existence of a Killing tensor with the separability of the KG equation seem to apply only in Einstein-Maxwell spaces, which our reduced 4-dimensional spacetime is not.  As a result of this, we don't expect separability of the KG equation to be possible for the black ring.  A straightforward calculation shows that this is indeed the case.  That is, taking an ansatz \[ \varphi(t,x,\phi,y,\psi) = e^{-i \Phi \phi} e^{-i \Psi \psi} X(x) Y(y) \] does not render the massless 5-dimensional KG equation $\Box \varphi = 0$ into a separable form.

\section{Discussion and Outlook} \label{sec:disc}
In this paper we have studied several aspects of the doubly-spinning black ring and noted that, although the metric is at first glance very complicated, it is possible to make progress in studying its properties analytically.  We have seen that in some senses the doubly-spinning system is more complicated, and richer, than the singly-spinning one, while other properties remain largely similar.  A brief summary of the original results in this paper was given in the introduction.

Some interesting questions remain.  We have not analysed in detail the paths of the axis geodesics in this paper, since doing so is very complicated, but it might be interesting to do this and see if any new behaviour occurs that does not appear in the singly spinning case.  These results could perhaps be useful in calculations of scattering cross sections; Gooding and Frolov have recently studied this problem in the Myers-Perry case \cite{Gooding:2008}.

As yet, there has been no attempt in the literature to construct the maximal analytic extension of a doubly-spinning black ring.  The coordinates of Section \ref{sec:coord} could potentially be useful in doing this; as they are the first example in the literature of an explicit coordinate system covering the horizon.

We have investigated possible links between our results, and the class of metrics described by \cite{Krtous:2008}, \cite{Houri:2008}.  We have found that the 4-dimensional spacetime obtained by dimensional reduction along $\pd/\pd t$ in the ergoregion is conformal to a metric falling into this class, if, and only if, the black ring is singly spinning.  This provides a qualitative, algebraic difference between the singly spinning and doubly spinning cases.

We might ask whether the unbalanced version \cite{Morisawa} of the ring has similar properties.  Studying the most general form of the unbalanced metric would be difficult, as it is extremely complicated, but some progress on this question can be made by looking at the limit where the black ring has rotation only in the $S^2$ direction, as derived by Figueras \cite{Figueras:2005}.  It turns out that here, no separation of the HJ equation is possible in ring-like coordinates; so this separability, and possibly the conformal Killing tensor structure associated with it, may rely in some way on the balancing condition being satisfied.  However, in the unbalanced, singly-spinning case \cite{ER:2006}, separation is possible, so the exact nature of this relationship is unclear.

\subsection*{Acknowledgements}
I would like to thank my supervisor, Harvey Reall for initially suggesting this project, as well as providing lots of useful advice and suggestions along the way.  I'd also like to thank Hari Kunduri for useful discussions about various aspects of the paper, Mukund Rangamani for sharing his work on ergoregions (see \cite{Elvang:2008erg}), and David Kubiz\v n\' ak, Tsuyoshi Houri and Yukinori Yasui for useful discussions on the role of conformal Killing and Killing-Yano tensors in studying higher-dimensional spacetimes.  I am funded by STFC.

\appendix
\section{Construction of an exact solution to (\ref{eqn:tdot})} \label{app:tdot}
In Section \ref{sec:coord} we looked to construct a solution to (\ref{eqn:tdot}) in order to construct a new time coordinate that is constant along one of our zero energy, null geodesics.  Here we derive an exact solution explicitly, mainly to demonstrate why using it is not the most convenient option for getting a new set of coordinates.

Note that, assuming that our separable solutions for $\eta^\phi$ and $\eta^\psi$ are the correct ones we can rewrite (\ref{eqn:tdot}) as \begin{equation} \label{eqn:chpde} f(x) \frac{\pd\eta}{\pd x} + g(y) \frac{\pd\eta}{\pd y} = h(x,y) \end{equation} where \begin{eqnarray*} f(x) &=& \pm \sqrt{\xi(x)} \\
g(y) &=& -\sqrt{\zeta(y)} \\
h(x,y) &=& \frac{\Omf (-\Axy \Phi + \Lxy \Psi) + \Omy (\Lxy \Phi + \Ayx \Psi)}{\Gx \Gy} . \end{eqnarray*}

To find a solution to this, we apply the method of characteristics.  Note that the characteristic curves follow the same paths in the $x-y$ plane as the geodesics, with the parameter $s$ a non-affine parameter along them.  We pick an arbitrary initial surface $y=b$, and pick our initial data to be $\eta(x,b)=0$.  The non-characteristic condition for surfaces of constant $y$ is that $g(y)\neq 0$.  This fails at $y=y_0$, so we must pick $b<y_0$, and clearly the initial surface should also lie outside the horizon.  Thus, we are free to choose any arbitrary $b$ with $ y_h < b < y_0$.  The initial surface can be parametrised as $\{(a,b)\}_{a\in [-1,1]}$, and given this the characteristic curves $(x(s;a),y(s;a))$ obey the equations \[ \frac{dx}{ds}(s;a) = f(x(s;a)) \eqand \frac{dx}{ds}(s;a) = g(y(s;a)), \] with solutions given implicitly by \[ \int_a^{x(s;a)} \frac{dx'}{f(x')} = s \; \; \mathrm{and} \;\; \int_b^{y(s;a)} \frac{dy'}{g(y')} = s . \]  Now define\footnote{This definition implicitly assumes the $x$ motion to be in the positive direction, the argument runs through in basically the same way with the opposite choice of sign.} \[ F: [-1,1] \rightarrow \left[ 0, \int_{-1}^1 \frac{dx'}{f(x')} \right] \eqand \Gamma : [y_h, b] \rightarrow \left[ 0, \int_{y_h}^b \frac{dy'}{\sqrt{\zeta(y')}} \right]  \]  by  \[ F(x)  \equiv \int_{-1}^x \frac{dx'}{f(x')} = \pm \int_{-1}^x \frac{dx'}{\sqrt{\xi(x)}} \eqand \Gamma(y) \equiv \int_b^y \frac{dy'}{g(y')} = \int_y^b \frac{dy'}{\sqrt{\zeta(y')}}. \]  Note that both $F$ and $\Gamma$ are bijective, and hence have well defined inverses.  Therefore, we can write \begin{equation} x(s;a) = F^{-1} (s+F(a)) \eqand y(s;a) = \Gamma^{-1} (s) .\label{eqn:flowmap} \end{equation}  Now, by (\ref{eqn:chpde}), \[ \frac{d}{ds} \eta(x(s;a),y(s;a)) = h(x(s;a),y(s;a)) \] and integrating this gives \[ \eta(x(s),y(s)) = \eta(a,b) + \int_0^s h\left[ F^{-1} (s' + F(a)), \Gamma^{-1} (s')\right] ds'. \]  Finally, we invert (\ref{eqn:flowmap}), change variables $ds' = d(\Gamma(y'))$ in the integral and insert our initial data $\eta(a,b)=0$ to give \begin{equation} \label{eqn:etat} \eta(x,y) = \int_b^y \frac{ h\left[ F^{-1}\left(\Gamma(y')-\Gamma(y) +F(x) \right), y' \right]}{g(y')} dy'. \end{equation}  This is a well defined solution to the system, which reduces to the known solution for the singly spinning case if we set $\nu=0$ (which means $h(x,y)$ is a function of $y$ only).  Unfortunately, it is not of a form where it is particularly convenient for use in a coordinate system.

It appears in the transformed metric via \begin{multline} dt + \Omf d\phi + \Omy d\psi = dv + \Omf d\tilde{\phi} + \Omy d\tilde{\psi} + \left( \frac{\pd \eta}{dx} + \Omf \frac{\pd \eta^\phi}{dx}  + \Omy \frac{\pd \eta^\psi}{dx} \right) dx \\ + \left( \frac{\pd \eta}{dy} + \Omf \frac{\pd \eta^\phi}{dy}  + \Omy \frac{\pd \eta^\psi}{dy} \right) dy \equiv dv + \tilde{\Omega}, \end{multline} where this final equality defines \[\tilde{\Om} =\Omf d\tilde{\phi} + \Omy d\tilde{\psi} + \tilde{\Om}_x dx + \tilde{\Om}_y dy . \]  Given our solution (\ref{eqn:etat}), we can write \[ \frac{ \pd \eta}{\pd x} = \frac{1}{f(x)} \int_b^y \frac{ (\pd_1 h)(x',y') f(x')}{g(y')} dy' \eqand \frac{ \pd \eta}{\pd y} = \frac{ h(x,y)}{g(y)} - \frac{1}{g(y)} \int_b^y \frac{ (\pd_1 h)(x',y') f(x')}{g(y')} dy', \] where \[ x'(y';x,y) \equiv F^{-1}\left(\Gamma(y')-\Gamma(y) +F(x) \right). \] Thus, \begin{eqnarray} \tilde{\Om}_x &=& \frac{\pd \eta}{\pd x} + \Omf \frac{\pd \eta^\phi}{\pd x} + \Omy \frac{\pd \eta^\psi}{\pd x} \nonumber \\
&=& \pm \frac{1}{ \sqrt{\tzx}} \left[ \frac{\Omf \tx +  \Omy \cx }{\Gx} + \int_b^y \frac{ (\pd_1 h)(x',y') f(x')}{g(y')} dy' \right] \\
\tilde{\Om}_y &=& \frac{\pd \eta}{\pd y} + \Omf \frac{\pd \eta^\phi}{\pd y} + \Omy \frac{\pd \eta^\psi}{\pd y} \nonumber \\
&=& \frac{1}{ \sqrt{\zy}} \left[ \frac{\Omf \tx +  \Omy \cx }{\Gx} - \int_b^y \frac{ (\pd_1 h)(x',y') f(x')}{g(y')} dy' \right] \\
\end{eqnarray} where \[ \theta(x) \equiv \beta(x) \Phi + \gamma(x) \Psi \eqand \chi(x) \equiv  \gamma(x) \Phi - \alpha(x) \Psi . \]

This form can then be inserted into the new metric (\ref{eqn:newmet}).  Note that we have not proved that this exact solution renders the metric regular at the horizon, and in fact it is not clear that it has this property.  The complicated form of the metric that we end up with here motivates us to look instead to merely solve the finiteness condition described above for the change of coordinates.

\section{Proof of Lemma 1} \label{app:symproof}
There are three parts to this lemma, the first two of which are essentially trivial.  Property 1 follows directly from the conformal Killing equation for $K'$, and it is easy to verify that $K$, and by extension $K'$ cannot be constructed from the metric and Killing vectors and is therefore independent of the metric.  Each independent CK tensor defines a conserved quantity $K^{a_1...a_p} p_{a_1}...p_{a_p}$, along a geodesic with null momentum $p_a$.  We already have 3 of these conserved quantities from $\pd/\pd\phi$, $\pd/\pd\psi$, and the metric itself.  In a 4-dimensional geometry, finding the geodesics reduces to solving 4 coupled first order ODEs, so there are only 4 independent conserved quantities.  If there was another tensor that we could add to $K$ to give a more general conformal Killing tensor, then this would itself give a new independent CK tensor, and hence a new conserved quantity, which is a contradiction.  It remains, therefore, to establish the non-trivial third property; the non-existence of a `square-root' of $K'$.

The equations for the components $K'^{xx}$, $K'^{yy}$, $K'^{x\phi}$, $K'^{x\psi}$, $K'^{y\phi}$, $K'^{y\psi}$ respectively of (\ref{eqn:fsquare}) can be written in the form \begin{eqnarray}
\label{eqn:Kxx} \frac{\Gx (1+C) }{\La^2} &=& \left(\! f^{x\phi} \; f^{x\psi} \! \right) \mathbf{M} \left( \! \begin{array}{c} f^{x\phi} \\ f^{x\psi}\end{array} \!\right) - \frac{ (f^{xy})^2}{\Gy} \\
\label{eqn:Kyy} \frac{\Gy (1-C)}{\La^2} &=& \left(\! f^{y\phi} \; f^{y\psi} \! \right) \mathbf{M} \left( \! \begin{array}{c} f^{y\phi} \\ f^{y\psi}\end{array} \!\right)+ \frac{ (f^{xy})^2}{\Gx} \\
\label{eqn:Kxp} f^{\phi\psi} \mathbf{M} \left( \!\begin{array}{c} f^{x\phi} \\ f^{x\psi} \end{array} \!\right) &=& \frac{f^{xy}}{\Gy} \left(\!\!\begin{array}{c} f^{y\psi} \\ - f^{y\phi}\end{array} \!\! \right) \\
\label{eqn:Kyp} f^{\phi\psi} \mathbf{M} \left( \!\begin{array}{c} f^{y\phi} \\ f^{y\psi} \end{array} \!\right) &=& \frac{f^{xy}}{\Gx} \left( \!\! \begin{array}{c} f^{x\psi} \\ - f^{x\phi}\end{array} \!\! \right)
\end{eqnarray} where \[ \mathbf{M} \equiv \frac{1}{\La^2} \left(\! \begin{array}{cc} h_{\phi\phi} & h_{\phi\psi}\\ h_{\phi\psi} & h_{\psi\psi} \end{array} \! \right)  = \frac{1}{\Hxy \Hyx} \left( \! \begin{array}{cc} \Ayx & -\Lxy \\ -\Lxy & -\Axy \end{array} \! \right) . \]

Contracting (\ref{eqn:Kxp}) with $G(y) \left( f^{x\phi} \; f^{x\psi}\right)$ and (\ref{eqn:Kxp}) with $G(x) \left( f^{y\phi} \; f^{y\psi}\right)$ gives us two new expressions for the LHS of equations (\ref{eqn:Kxx}), (\ref{eqn:Kyy}).  Substituting these in, and taking the difference of the resulting equations leaves us with \[ \frac{2 f^{\phi\psi} \Gx \Gy}{\La^2} = 0 \Rightarrow f^{\phi\psi} = 0. \]  Inserting this back into (\ref{eqn:Kxp}), (\ref{eqn:Kyp}) gives \[ f^{xy} \left( \!\begin{array}{c} f^{y\psi} \\ - f^{y\phi}\end{array} \!\right) = 0 =  f^{xy} \left( \!\begin{array}{c} f^{x\psi} \\ - f^{x\phi}\end{array} \!\right), \] and hence we must have $f^{xy}=0$ (since otherwise we would have all other components vanishing, which leads us into an immediate contradiction).

Given these results, we then consider the components $K'^{\phi\phi}$, $K'^{\psi\psi}$ and $K'^{\phi\psi}$:  \begin{eqnarray} 
\frac{1}{(1-\nu)^2} \left( \frac{\bxb}{\Gx} - \frac{\ayb}{\Gy} + p \right)  &=& \La^2 \left( \frac{(f^{x\phi})^2}{\Gx} - \frac{(f^{y\phi})^2}{\Gy} \right), \\
\frac{1}{(1-\nu)^2} \left( \frac{\byb}{\Gy} - \frac{\axb}{\Gx} + r \right)  &=& \La^2 \left( \frac{(f^{x\psi})^2}{\Gx} - \frac{(f^{y\psi})^2}{\Gy} \right) , \\
\frac{1}{(1-\nu)^2} \left( \frac{\gx}{\Gx} + \frac{\gy}{\Gy} + q \right) &=& \La^2 \left( \frac{f^{x\phi} f^{x\psi}}{\Gx} - \frac{f^{y\phi} f^{y\psi}}{\Gy} \right). \end{eqnarray}   We can use these three equations to express $(f^{y\phi})^2$, $(f^{y\psi})^2$ and $f^{y\phi} f^{y\psi}$ in terms of $(f^{x\phi})^2$, $(f^{x\psi})^2$ and $f^{x\phi} f^{x\psi}$, and then put this into (\ref{eqn:Kyy}).  Comparing this to (\ref{eqn:Kxx}) leads to a consistency condition \[ \frac{\ayb \byb + \gy^2 + (-p\byb + 2q \gy + r \ayb)\Gy}{\Gy^2} = \frac{\axb \bxb + \gx^2 + (p\axb + 2q \gx - r \bxb)\Gx}{\Gx^2}  \] that is independent of $C$.  This separates $x$ and $y$, and hence can only be satisfied if both sides are constant for some choice of constants $p$,$q$,$r$.  In the singly spinning case this holds since $\alpha(\xi) = 0 = \gamma(\xi)$ for all $\xi\in (-\infty, 1]$, and we can then choose $p=r=0$ to make both sides vanish.  In the doubly spinning case, however, we are required to set \[r=\lim_{x\rightarrow \pm 1} \frac{\axb}{\Gx} \] to avoid a pole in the RHS at $x=\pm 1$.  But these two limits are not the same for $\nu>0$, so we have a contradiction, which completes the proof of the Lemma.

\bibliographystyle{h-physrev}
\bibliography{ringgeo}

\begin{thebibliography}{10}

\bibitem{ER:2001}
R.~{Emparan} and H.~S. {Reall},
\newblock Phys. Rev. Lett. {\bf 88}, 101101 (2002), arXiv:hep-th/0110260.

\bibitem{HawkingEllis}
S.~W. {Hawking} and G.~F.~R. {Ellis},
\newblock {\em {The Large Scale Structure of Space-time}} (Cambridge University
  Press, 1973).

\bibitem{Galloway:2005}
G.~J. Galloway and R.~Schoen,
\newblock Commun. Math. Phys. {\bf 266}, 571 (2006), arXiv:gr-qc/0509107.

\bibitem{Myers:1986}
R.~C. {Myers} and M.~J. {Perry},
\newblock Ann. Phys. {\bf 172}, 304 (1986).

\bibitem{Elvang:2007bi}
H.~{Elvang} and M.~J. {Rodriguez},
\newblock JHEP {\bf 04}, 045 (2008), arXiv:0712.2425 [hep-th].

\bibitem{Elvang:2007sat}
H.~{Elvang} and P.~{Figueras},
\newblock JHEP {\bf 05}, 050 (2007), arXiv:hep-th/0701035.

\bibitem{Iguchi:2007}
H.~{Iguchi} and T.~{Mishima},
\newblock Phys. Rev. {\bf D75}, 064018 (2007), arXiv:hep-th/0701043.

\bibitem{Evslin:2007}
J.~Evslin and C.~Krishnan,
\newblock (2007), arXiv:0706.1231 [hep-th].

\bibitem{ER:2008}
R.~{Emparan} and H.~S. {Reall},
\newblock Living Reviews in Relativity  (2008), arXiv:0801.3471 [hep-th].

\bibitem{Pomeransky}
A.~A. {Pomeransky} and R.~A. {Sen'kov},
\newblock (2006), arXiv:hep-th/0612005.

\bibitem{Kunduri:2007}
H.~K. {Kunduri}, J.~{Lucietti}, and H.~S. {Reall},
\newblock Class. Quant. Grav. {\bf 24}, 4169 (2007), arXiv:0705.4214 [hep-th].

\bibitem{Morisawa}
Y.~{Morisawa}, S.~{Tomizawa}, and Y.~{Yasui},
\newblock Phys. Rev. {\bf D77}, 064019 (2008), arXiv:0710.4600 [hep-th].

\bibitem{ER:2006}
R.~{Emparan} and H.~S. {Reall},
\newblock Class. Quant. Grav. {\bf 23}, R169 (2006), arXiv:hep-th/0608012.

\bibitem{Elvang:2008erg}
H.~{Elvang}, P.~{Figueras}, G.~T. {Horowitz}, V.~E. {Hubeny}, and
  M.~{Rangamani},
\newblock (2008), arXiv:0810.2778 [gr-qc].

\bibitem{Hoskisson}
J.~{Hoskisson},
\newblock Phys. Rev. {\bf D78}, 064039 (2008), arXiv:0705.0117 [hep-th].

\bibitem{Chrusciel:2008}
P.~T. {Chrusciel} and J.~{Cortier},
\newblock (2008), arXiv:0807.2309 [gr-qc].

\bibitem{Walker:1970}
M.~{Walker} and R.~{Penrose},
\newblock Commun. Math. Phys. {\bf 18}, 265 (1970).

\bibitem{Chow:2008}
D.~D.~K. {Chow},
\newblock (2008), arXiv:0808.2728 [hep-th].

\bibitem{Hioki:2008}
K.~{Hioki} and U.~{Miyamoto},
\newblock Phys. Rev. {\bf D78}, 044007 (2008), arXiv:0805.3146 [gr-qc].

\bibitem{Chamblin:1996}
A.~{Chamblin} and R.~Emparan,
\newblock Phys. Rev. {\bf D55}, 754 (1997), hep-th/9607236.

\bibitem{Benenti:1979}
S.~{Benenti} and M.~{Francaviglia},
\newblock Gen. Rel. and Grav. {\bf 10}, 79 (1979).

\bibitem{Krtous:2008}
P.~Krtous, V.~P. Frolov, and D.~Kubiznak,
\newblock (2008), arXiv:0804.4705 [hep-th].

\bibitem{Houri:2008}
T.~Houri, T.~Oota, and Y.~Yasui,
\newblock (2008), arXiv:0805.3877 [hep-th].

\bibitem{Chen:2006}
W.~Chen, H.~Lu, and C.~N. Pope,
\newblock Class. Quant. Grav. {\bf 23}, 5323 (2006), arXiv:hep-th/0604125.

\bibitem{Gooding:2008}
C.~Gooding and A.~V. Frolov,
\newblock Phys. Rev. {\bf D77}, 104026 (2008), arXiv:0803.1031 [gr-qc].

\bibitem{Figueras:2005}
P.~{Figueras},
\newblock JHEP {\bf 07}, 039 (2005), hep-th/0505244.

\end{thebibliography}
\end{document}